\documentclass[aps,twocolumn,pre,amsmath,amssymb,showpacs,notitlepage,superscriptaddress]{revtex4-1}

\usepackage[utf8]{inputenc}
\usepackage[T1]{fontenc}
\usepackage{textcomp}
\usepackage[english]{babel}
\usepackage{color}
\usepackage{amssymb}
\usepackage{amsmath}
\usepackage{amsfonts}
\usepackage{amsopn}
\usepackage{amsbsy}
\usepackage{enumerate}
\usepackage[plainpages=false,pdfpagelabels,breaklinks=true,pdfborderstyle={/S/U/W 1.2}]{hyperref}
\usepackage{graphicx}
\usepackage{units}
\usepackage[cal=boondoxo,bb=boondox]{mathalfa}
\usepackage{dsfont}
\usepackage{tabularx}
\usepackage[activate={true,nocompatibility},final,tracking=true,kerning=true,spacing=true,stretch=10,shrink=10]{microtype}

\usepackage[scaled]{beramono}
\usepackage{helvet}
\usepackage[charter]{mathdesign} 
\SetMathAlphabet{\mathcal}{normal}{OMS}{cmsy}{m}{n} 
\SetMathAlphabet{\mathcal}{bold}{OMS}{cmsy}{m}{n} 
\providecommand{\ie}{i.e.~}
\providecommand{\eg}{e.g.~}
\providecommand{\cf}{cf.~}

\providecommand{\R}{\mathbb{R}}

\providecommand{\C}{\mathbb{C}}
\renewcommand{\C}{\mathbb{C}}

\providecommand{\ii}{\mathrm{i}}
\providecommand{\e}{\mathrm{e}}
\renewcommand{\Re}{\mathrm{Re} \,}

\providecommand{\Hil}{\mathcal{H}}

\providecommand{\eps}{\varepsilon}

\providecommand{\dd}{\mathrm{d}}
\providecommand{\id}{\mathds{1}}

\providecommand{\order}{\mathcal{O}}
\providecommand{\Fourier}{\mathcal{F}}

\providecommand{\abs}[1]{\left \lvert #1 \right \rvert}
\providecommand{\sabs}[1]{\lvert #1 \vert}

\providecommand{\norm}[1]{\left \lVert #1 \right \rVert}

\providecommand{\scpro}[2]{\left \langle #1 , #2 \right \rangle}
\providecommand{\sscpro}[2]{\langle #1 , #2 \rangle}
\providecommand{\bscpro}[2]{\bigl \langle #1 , #2 \bigr \rangle}

\providecommand{\sket}[1]{\vert #1 \rangle}

\providecommand{\sbra}[1]{\langle #1 \vert}

\providecommand{\Krein}{\mathcal{K}}

\providecommand{\WS}{M}
\providecommand{\BZ}{M^*}

\usepackage{diagxy}

\begin{document}

\title{The Krein-Schrödinger Formalism of Bosonic BdG and \\ Certain Classical Systems and Their Topological Classification}

\author{Max Lein}
\email{max.lein@tohoku.ac.jp}
\affiliation{Advanced Institute of Materials Research, Tohoku University, Sendai 980-8577, Japan}

\author{Koji Sato}
\email{koji.sato@imr.tohoku.ac.jp}
\affiliation{Institute for Materials Research, Tohoku University, Sendai 980-8577, Japan}

\begin{abstract}
	To understand recent works on classical and quantum spin equations and their topological classification, we develop a unified mathematical framework for bosonic BdG systems and associated classical wave equations; it applies not just to equations that describe quantized spin excitations in magnonic crystals but more broadly to other systems that are described by a BdG hamiltonian. Because here the generator of dynamics, the analog of the hamiltonian, is para- aka Krein-hermitian but not hermitian, the theory of Krein spaces plays a crucial role. For systems which are thermodynamically stable, the classical equations can be expressed as a “Schrödinger equation” with a hermitian “hamiltonian”. 
	We then proceed to apply the Cartan-Altland-Zirnbauer classification scheme: to properly understand what topological class these equations belong to, we need to conceptually distinguish between symmetries and constraints. Complex conjugation enters as a particle-hole \emph{constraint} (as opposed to a symmetry), since classical waves are necessarily real-valued. 
	Because of this distinction only commuting symmetries enter in the topological classification. Our arguments show that the equations for spin waves in magnonic crystals are a system of class~A, the same topological class as quantum hamiltonians describing the Integer Quantum Hall Effect. Consequently, the magnonic edge modes first predicted by Shindou et al.\ \cite{Shindou_et_al:chiral_magnonic_edge_modes:2013} are indeed analogs of the Quantum Hall Effect, and their net number is topologically protected.
\end{abstract}

\maketitle

\section{Introduction} 
\label{intro}
Shindou et al.\ proposed in \cite{Shindou_et_al:chiral_magnonic_edge_modes:2013} that topologically protected edge modes should exist in periodic media for spin waves, so-called magnonic crystals. The basis of their work was Haldane's insight \cite{Raghu_Haldane:quantum_Hall_effect_photonic_crystals:2008} that topological effects are not particular to quantum systems, but \emph{bona fide} wave effects; more specifically, Haldane proposed that the existence of boundary modes and their net number are related to a topological invariant, the Chern number. His prediction was confirmed in a number of spectacular experiments with various classical waves, 
including electromagnetic waves \cite{Wang_et_al:edge_modes_photonic_crystal:2008,Rechtsman_Zeuner_et_al:photonic_topological_insulators:2013,DeNittis_Lein:symmetries_electromagnetism:2017}, certain acoustic waves \cite{Fleury_et_al:breaking_TR_acoustic_waves:2014,Safavi-Naeini_et_al:2d_phonoic_photonic_band_gap_crystal_cavity:2014,Peano_Brendel_Schmidt_Marquardt:topological_phases_sound_light:2015,Chen_Zhao_Mei_Wu:acoustic_frequency_filter_topological_phononic_crystals:2017} and coupled pendula \cite{Suesstrunk_Huber:mechanical_topological_insulator:2015,Suesstrunk_Huber:classification_mechanical_metamaterials:2016}. 

By the same token, Shindou et al.\ proposed that Haldane's photonic bulk-edge correspondence also applies to magnonic crystals, where the net number of boundary modes is given by the Chern number 
\begin{align}
	C = \frac{1}{2\pi} \int_{\BZ} \dd k \; \mathrm{Tr} \, \bigl ( \Omega(k) \bigr )
	\label{intro:eqn:Chern_number}
\end{align}
that is computed as the Brillouin zone average of the Berry curvature 
\begin{align}
	\Omega_{jn}(k) = \partial_{k_j} \mathcal{A}_n(k) - \partial_{k_n} \mathcal{A}_j(k) 
	\label{intro:eqn:Berry_curvature}
\end{align}
which in turn is derived from the Berry connection 
\begin{align}
	\mathcal{A}_{\alpha \beta \, n}(k) = \ii \; \bscpro{\varphi_{\alpha}(k) \, }{ \, \sigma_3 \, \partial_{k_n} \varphi_{\beta}(k)} 
	. 
	\label{intro:eqn:Berry_connection}
\end{align}
At first glance, everything \emph{seems} identical to the Quantum Hall Effect or its photonic analog, but there is a critical difference to the other cases: the inner product used to compute Berry connection and curvature, 
\begin{align}
	\scpro{\Phi}{\Psi}_{\sigma_3} = \bscpro{\Phi}{\sigma_3 \, \Psi}
	, 
	\label{intro:eqn:Krein_inner_product}
\end{align}
is indeterminate and therefore is \emph{not} a scalar product. And because $\scpro{\Psi}{\Psi}_{\sigma_3}$ can be positive, negative or even zero when $\Psi \neq 0$, it is \emph{a priori} not clear whether Eq.~\eqref{intro:eqn:Berry_connection} is really a connection and Eq.~\eqref{intro:eqn:Berry_curvature} really a curvature. Consequently, if $\Omega$ is not a curvature, then Eq.~\eqref{intro:eqn:Chern_number} does not even need to be an integer, let alone a label of a topological phase. 

Understanding Shindou et al.'s work from first principles motivated us to look at bosonic BdG systems and related classical systems more broadly as they all share the same mathematical structure. On the \emph{quantum} side the standard starting point is to consider the equations of motion of creation and annihilation operators 
\begin{align}
	\ii \frac{\partial}{\partial t} \left (
	\begin{matrix}
		a(t,k) \\
		a^{\dagger}(t,-k) \\
	\end{matrix}
	\right ) = \sigma_3 \, H_{\mathrm{sp}}(k) \left (
	\begin{matrix}
		a(t,k) \\
		a^{\dagger}(t,-k) \\
	\end{matrix}
	\right )
	\label{intro:eqn:second_quantized_spin_equations}
\end{align}
that are governed by a single-particle BdG hamiltonian $H_{\mathrm{sp}}(k)$. This operator is usually assumed to be “thermodynamically stable” (\cf Eq.~\eqref{Krein_space_formalism:eqn:stability_condition}) and comes furnished with an even particle-hole-type symmetry $C$, which ensures that time-evolved creation and annihilation operators are adjoints of one another; therefore, in the absence of other symmetries (chiral- or time-reversal-type symmetries) $H_{\mathrm{sp}}(k)$ is an operator of class~D \cite{Altland_Zirnbauer:superconductors_symmetries:1997,Chiu_Teo_Schnyder_Ryu:classification_topological_insulators:2016}. 

For all systems of interest $H_{\mathrm{sp}}$ is not block-diagonal and therefore the operator
\begin{align*}
	\sigma_3 \, H_{\mathrm{sp}} \neq H_{\mathrm{sp}} \, \sigma_3 = \bigl ( \sigma_3 \, H_{\mathrm{sp}} \bigr )^{\dagger}
\end{align*}
is \emph{not hermitian}.

There is a class of corresponding \emph{classical} wave equations, whose structure mimics Eq.~\eqref{intro:eqn:second_quantized_spin_equations}: the operator $W^{-1} \, D$ which generates the time evolution of the classical wave $m(t)$ via the “Schrödinger equation”
\begin{align}
	\ii \partial_t m(t) = W^{-1} \, D \, m(t) 
	,
	&&
	m(t_0) = n = \overline{n}
	,
	\label{intro:eqn:classical_wave_equation}
\end{align}
is the product of two hermitian operators $W = W^{\dagger}$ and $D = D^{\dagger}$; as the notation suggests, we assume that $W$ is invertible. What distinguishes the wave equations relevant here is the assumption that \emph{$W$ does not have a fixed sign} — just like $\sigma_3$ above. 

In fact, for a given bosonic BdG equation~\eqref{intro:eqn:second_quantized_spin_equations}, there is a corresponding classical equation where $W = \sigma_3$ and $D = H_{\mathrm{sp}}$. And properties of the quantum equation — including their topological classification — are immediately linked to their classical counterpart. Logically speaking, though, it is more consistent to view the quantum equation~\eqref{intro:eqn:second_quantized_spin_equations} as the second quantization of the classical wave equation~\eqref{intro:eqn:classical_wave_equation}. 

\emph{The main aim of our article is to derive a unified mathematical framework to treat systems described by Eq.~\eqref{intro:eqn:second_quantized_spin_equations} or \eqref{intro:eqn:classical_wave_equation}, and to obtain their topological classification.}
\medskip

\noindent
We will add two new insights to the discussion: the first is a careful \emph{conceptual distinction between symmetries and constraints}: complex conjugation should not be regarded as a symmetry of $H = W \, D$, but is a \emph{constraint on the relevant states}. 
Our analysis in Section~\ref{Krein_space_formalism} below shows that the physical system is \emph{not} in class~D (the topological class of $W \, D$), but rather in \emph{class~A} — the same topological class as quantum hamiltonians which model the Quantum Hall Effect. That is because the particle-hole-type “symmetry” represents complex conjugation of classical waves, and since classical waves are necessarily real-valued, complex conjugation enters as a \emph{constraint} rather than a symmetry that can be selectively broken. We have emphasized this point in earlier works on the Schrödinger formalism of classical waves \cite{DeNittis_Lein:Schroedinger_formalism_classical_waves:2017} and the topological classification of electromagnetic media \cite{DeNittis_Lein:symmetries_electromagnetism:2017}. 

However, our earlier works \cite{DeNittis_Lein:Schroedinger_formalism_classical_waves:2017,DeNittis_Lein:symmetries_electromagnetism:2017} do not directly apply to the classical wave equations studied here — the operator which governs the dynamics is not hermitian. Instead, this operator is Krein- or para-hermitian: if we endow the Hilbert space with the indeterminate Krein inner product $\scpro{\phi}{\psi}_W = \scpro{\phi}{W \, \psi}$, we obtain a so-called Krein space. While facets of the theory of Krein spaces have been used by physicists for a long time, it seems that very few works in theoretical physics have systematically exploited the mathematical tools from the theory of Krein spaces. 

For thermodynamically stable “hamiltonians”, where $D$ is positive definite, we are able to extend our arguments from \cite{DeNittis_Lein:Schroedinger_formalism_classical_waves:2017} to classical and second-quantized spin waves. The first step is to establish a one-to-one correspondence between the \emph{real} vector space of real-valued classical classical waves and the \emph{complex} vector space $\Hil_+(k)$ composed of complex positive frequency waves. On the subset of complex positive frequency waves the inner product~\eqref{intro:eqn:Krein_inner_product} is positive definite, and therefore a \emph{bona fide} scalar product; this makes $\Hil_+(k)$ into a Hilbert space. The positive definiteness of $\scpro{\phi}{\psi}_W = \bscpro{\phi}{W \, \psi}$ on that subspace \emph{a posteriori} justifies why Eq.~\eqref{intro:eqn:Berry_connection} is a curvature and Eq.~\eqref{intro:eqn:Berry_curvature} a curvature. While we initially analyze only classical wave equations, we prove in Section~\ref{second_quantized_equations} that the topological classification of the classical spin equations from Section~\ref{magnonic_crystal:bulk_edge_correspondence} carries over to the second-quantized equation~\eqref{intro:eqn:second_quantized_spin_equations}. 

The Schrödinger formalism for bosonic BdG systems allows us to single out physically meaningful symmetries in spin systems; for instance, we identify the time-reversal-type symmetries that are broken and give their physical interpretation. Importantly, it is not the magnetic field that breaks these time-reversal-type symmetries, but rather anisotropy in space (\cf Section~\ref{quantum_vs_classical:classical:physical_interpretation_TRS}).%
\medskip

\noindent
\textbf{Outline} 
We start by comparing the mathematical description and symmetries of the simplest possible quantum and classical spin equations in Section~\ref{quantum_vs_classical}; the goal is to outline the strategy we use to derive the Schrödinger formalism of a \emph{classical} wave equation. The insights gathered there are then applied to the linearized Landau-Lifshitz equations in Section~\ref{Krein_space_formalism}: after a short primer on the theory of Krein spaces, we discuss the spin “hamiltonian” of the complexified equations as an operator on Krein space and study its symmetries. Then we discard the superfluous degrees of freedom that stem from the complexification to arrive at the Schrödinger formulation of the classical linearized spin equations, which in turn allows us to apply the Cartan-Altland-Zirnbauer scheme. Up until this point we did not need to assume that the medium for the spins is periodic. However, when the spin travels in a magnonic crystal, we can make the topological bulk classification and bulk-boundary correspondence explicit (Section~\ref{magnonic_crystals}). Then we show that our arguments indeed apply to all bosonic systems described by a BdG hamiltonian that is thermodynamically stable (Section~\ref{other_systems}). Lastly, we explain how the topological classification extend from the classical to the second-quantized equations (Section~\ref{second_quantized_equations}). 

\section{Symmetries of classical and analogous quantum spin equations} 
\label{quantum_vs_classical}
The classification theory for topological insulators \cite{Altland_Zirnbauer:superconductors_symmetries:1997,Chiu_Teo_Schnyder_Ryu:classification_topological_insulators:2016,Ozawa_et_al:review_topological_photonics:2018} was first developed for quantum systems, and in order to adapt and apply these methods to spin systems we need to reformulate the dynamical equation in the form of a Schrödinger equation. By this we mean that we find a formulation of the classical equations in terms of a hermitian operator that acts on a \emph{complex} Hilbert space; this is only a \emph{mathematical procedure}, spin waves are still classical and 
expressing them in terms of complex waves has nothing to do with probabilities. 
This Schrödinger formalism allows us to identify the number of discrete symmetries and their nature that determine the topological class, which in turn allows us to list the topological invariants that are supported by the system. 

These ideas have been applied to other classical waves \cite{DeNittis_Lein:Schroedinger_formalism_classical_waves:2017,DeNittis_Lein:symmetries_electromagnetism:2017}, but it is helpful to illustrate the essential features with a very simple set of equations that structurally resemble those for magnons. To that end we will compare a quantum spin with a classical spin, both moving in a magnetic field whose dynamical equations appear to be identical.

\subsection{A quantum spin-$\nicefrac{1}{2}$ precessing in a magnetic field} 
\label{quantum_vs_classical:quantum}
The simplest quantum system is a quantum spin-$\nicefrac{1}{2}$ subjected to a constant magnetic field. If we choose a coordinate system so that the magnetic field points in the direction $e_2 = (0,1,0)^{\mathrm{T}}$, the relevant Schrödinger equation governing the dynamics is 
\begin{align}
	\ii \frac{\partial}{\partial t} \left (
	\begin{matrix}
		\psi_1(t) \\
		\psi_2(t) \\
	\end{matrix}
	\right ) &= \left (
	\begin{matrix}
		0 & - \ii \, \omega \\
	 	+ \ii \, \omega & 0 \\
	\end{matrix}
	\right ) \left (
	\begin{matrix}
		\psi_1(t) \\
		\psi_2(t) \\
	\end{matrix}
	\right )
	. 
	\label{quantum_vs_classical:eqn:quantum_spin_1_2_Schroedinger_equation}
\end{align}
Here, the frequency of precession $\omega = \frac{\hbar}{2} \, g H_0$ collects all physical constants and the spin state is described by $\Psi(t) = \bigl ( \psi_1(t) , \psi_2(t) \bigr )^{\mathrm{T}} \in \Hil_{\C} = \C^2$ that is normalized to $1$. The fact that the Hamiltonian $H = \omega \, \sigma_2 = H^{\dagger}$ is hermitian leads to conservation of probability. Note that the concept of hermitian operators necessitates the use of \emph{complex} (as opposed to real) Hilbert spaces. 

Due to the simplicity of the system, we can list all symmetries, three of them are given by the Pauli matrices and define (linear) unitaries, 
\begin{subequations}\label{quantum_vs_classical:eqn:Pauli}
	\begin{align}
		\sigma_{1,3} \, H \, \sigma_{1,3}^{-1} &= - H 
		, 
		\label{quantum_vs_classical:eqn:Pauli:chiral}
		\\
		(\ii \sigma_2) \, H \, (\ii \sigma_2)^{-1} &= + H 
		\label{quantum_vs_classical:eqn:Pauli:ordinary}
		, 
	\end{align}
\end{subequations}
whereas the four other are complex conjugation $C$, 
\begin{align}
	C \, H \, C = - H
	, 
	\label{quantum_vs_classical:eqn:complex_conjugation_particle_hole}
\end{align}
and products of the three Pauli matrices with complex conjugation,
\begin{subequations}\label{quantum_vs_classical:eqn:Pauli_C}
	\begin{align}
		(\sigma_{1,3} C) \, H \, (\sigma_{1,3} C)^{-1} &= + H 
		, 
		\label{quantum_vs_classical:eqn:Pauli_C:time_reversal}
		\\
		(\ii \sigma_2 C) \, H \, (\ii \sigma_2 C)^{-1} &= - H 
		\label{quantum_vs_classical:eqn:Pauli_C:particle_hole}
		. 
	\end{align}
\end{subequations}
The reason why we prefer to equivalently use the real, antihermitian matrix $\ii \sigma_2$ rather than $\sigma_2$ will become apparent in the next subsection. 

We can straightforwardly identify the nature of these symmetries in the context of classification theory for topological insulators \cite{Altland_Zirnbauer:superconductors_symmetries:1997,Chiu_Teo_Schnyder_Ryu:classification_topological_insulators:2016}. For example, the linear, anticommuting symmetries $\sigma_{1,3}$ (\cf Eq.~\eqref{quantum_vs_classical:eqn:Pauli:chiral}) are classified as \emph{chiral-type symmetries}, whereas the antilinear, commuting symmetries $\sigma_{1,3} C$ square to $+ \id$ and are \emph{even time-reversal-type symmetries}. The even particle-hole-type symmetry $C$ (\cf Eq.~\eqref{quantum_vs_classical:eqn:complex_conjugation_particle_hole}) will play an important role below. Table~\ref{quantum_vs_classical:table:quantum_symmetries} summarizes the classification of all 7 symmetries. 

\begin{table}
	\begin{center}
		\renewcommand{\arraystretch}{1.5}
		\newcolumntype{A}{>{\centering\arraybackslash\normalsize} m{25mm} }
		\begin{tabular}{A | c}
			\textbf{Symmetry} & \textbf{TI Classification} \\ \hline \hline
			$\sigma_1$ & chiral \\ \hline
			$\ii \sigma_2$ & ordinary \\ \hline
			$\sigma_3$ & chiral \\ \hline
			$C$ & +PH \\ \hline
			$\sigma_1 \, C$ & +TR \\ \hline
			$\ii \sigma_2 \, C$ & –PH \\ \hline
			$\sigma_3 \, C$ & +TR \\
		\end{tabular}
	\end{center}

	\caption{Summary of all symmetries of quantum spin system. Ordinary symmetries are those that are unitary and commute with the Hamiltonian. +TR and $\pm \mathrm{PH}$ are abbreviations for even time-reversal and even/odd particle-hole symmetries (depending on whether the symmetries square to $\pm \id$). }
	\label{quantum_vs_classical:table:quantum_symmetries}
\end{table}
%

\subsection{A classical precessing spin in $\R^3$} 
\label{quantum_vs_classical:classical}
The equations that describe precession of a classical spin with frequency $\omega$ in a magnetic field pointing in the $e_3$-direction, 
\begin{align*}
	\frac{\partial }{\partial t} \left (
	\begin{matrix}
		M_1(t) \\
		M_2(t) \\
		M_3(t) \\
	\end{matrix}
	\right ) &= \left (
	\begin{matrix}
		0 \\
		0 \\
		\omega \\
	\end{matrix}
	\right ) \times \left (
	\begin{matrix}
		M_1(t) \\
		M_2(t) \\
		M_3(t) \\
	\end{matrix}
	\right )
	, 
\end{align*}
can be brought into the form of \eqref{quantum_vs_classical:eqn:quantum_spin_1_2_Schroedinger_equation}: exploiting that $M_3(t) = \mathrm{const}.$ is a conserved quantity, we drop this component from the equations and multiply both sides by the imaginary unit $\ii$ to obtain 
\begin{align}
	\ii \frac{\partial }{\partial t} \left (
	\begin{matrix}
		M_1(t) \\
		M_2(t) \\
	\end{matrix}
	\right ) &= \left (
	\begin{matrix}
		0 & - \ii \, \omega \\
		+ \ii \, \omega & 0 \\
	\end{matrix}
	\right ) \left (
	\begin{matrix}
		M_1(t) \\
		M_2(t) \\
	\end{matrix}
	\right )
	. 
	\label{quantum_vs_classical:eqn:classical_real_spin_equation}
\end{align}
Compared to Eq.~\eqref{quantum_vs_classical:eqn:quantum_spin_1_2_Schroedinger_equation} this equation is \emph{only} defined for \emph{real} spin vectors $M(t) = \bigl ( M_1(t) , M_2(t) \bigr )^{\mathrm{T}} \in \Hil_{\R} = \R^2$. Instead of total probability, the conserved quantity $\abs{M(t)} = \mathrm{const}.$ that follows from the orthogonality of the evolution is now spin length. 

On the real Hilbert space $\Hil_{\R} = \R^2$ all (orthogonal) symmetry operators (matrices), that map real vectors onto real vectors, can be written as a linear combination of 
\begin{subequations}\label{quantum_vs_classical:eqn:real_symmetries}
	\begin{align}
		V_{1,3}^{\R} &= \sigma_{1,3} 
		, 
		\label{quantum_vs_classical:eqn:real_symmetries:1_3}
		\\
		V_2^{\R} &= \ii \sigma_2 
		. 
		\label{quantum_vs_classical:eqn:real_symmetries:2}
	\end{align}
\end{subequations}
This also explains why we preferred to use the \emph{real} matrix $\ii \sigma_2$ in the previous section rather than $\sigma_2$, the latter is not a well-defined operator on $\Hil_{\R} = \R^2$. 

However, for a number of reasons — including understanding the topological origin of boundary modes in magnonic crystals — it is useful to consider also the classical equations on complex Hilbert spaces.

\subsubsection{Complexifying the equations of motion} 
\label{quantum_vs_classical:classical:complexification}
To solve Eq.~\eqref{quantum_vs_classical:eqn:classical_real_spin_equation} efficiently, we need to work with complex vectors. The default approach is to express real-valued solutions 
\begin{align}
	M(t) &= \left (
	\begin{matrix}
		\cos \omega t & - \sin \omega t \\
		\sin \omega t & \cos \omega t \\
	\end{matrix}
	\right ) \left (
	\begin{matrix}
		a \\
		b \\
	\end{matrix}
	\right )
	\label{quantum_vs_classical:eqn:real_solution_express_via_complex_waves}
	\\
	&= \tfrac{1}{2} (a - \ii b) \, \e^{- \ii \omega t} \, \left (
	\begin{smallmatrix}
		1 \\
		+ \ii \\
	\end{smallmatrix}
	\right ) + \tfrac{1}{2} (a + \ii b) \, \e^{+ \ii \omega t} \, \left (
	\begin{smallmatrix}
		1 \\
		- \ii \\
	\end{smallmatrix}
	\right )
	\notag 
	\\
	&= \psi_+(t) + \psi_-(t) 
	\notag 
\end{align}
as a linear combination of complex eigenstates of frequency $\pm \omega$. While mathematically we consider \eqref{quantum_vs_classical:eqn:classical_real_spin_equation} on $\Hil_{\C} = \C^2$, the \emph{same} complex Hilbert space as in the quantum case, the crucial difference is that only real-valued spins $M \in \Hil_{\R} = \R^2 \subset \C^2 = \Hil_{\C}$ have physical significance. \emph{States with non-vanishing imaginary part have no physical meaning,} they are an artifact of doubling the degrees of freedom for convenience's sake. 

While it may seem that working on $\Hil_{\C}$ brings all 7 “quantum symmetries”~\eqref{quantum_vs_classical:eqn:Pauli}–\eqref{quantum_vs_classical:eqn:Pauli_C} into play, this is \emph{not} the case: complex conjugation acts trivially on real-valued spins, $C M = M \in \Hil_{\R} = \R^2$, so for \emph{physical} states complex conjugation is \emph{not a meaningful symmetry}. That is why we prefer to call $C$ a \emph{constraint} rather than an “even particle-hole-type \emph{symmetry}”. Indeed, $C \, H \, C = - H$ implies that the time evolution 
\begin{align*}
	\e^{- \ii t H} \, \Re = \Re \, \e^{- \ii t H} 
\end{align*}
commutes with the real part operator $\Re = \tfrac{1}{2} (\id + C)$ and therefore maps real initial states onto real-valued solutions. So if $H$ is supposed to describe a \emph{classical} spin or wave, \emph{then an “unbreakable” even particle-hole-type constraint is baked into the complexified equations}; it arises from the real-valuedness of the classical waves. Consequently, the actions of $\sigma_{1,3}$ and $\sigma_{1,3} C$ coincide on the subspace $\Hil_{\R} = \R^2$ of physical states. Thus, the number of distinct physical symmetries reduces from 7 to 3. 

This ambiguity causes problems when we want to adapt the Cartan-Altland-Zirnbauer (CAZ) classification scheme \cite{Altland_Zirnbauer:superconductors_symmetries:1997,Chiu_Teo_Schnyder_Ryu:classification_topological_insulators:2016} for topological insulators. We are \emph{seemingly} able to choose between \eg using the \emph{chiral-type symmetries} $\sigma_{1,3}$ or the \emph{even time-reversal-type symmetries} $\sigma_{1,3} \, C$ for our topological classification. So it is unclear whether we are considering a system of class~AIII or class~AI. If we added complex conjugation as a separate symmetry, the classification would be different still, namely class~BDI. We will explain later in Section~\ref{Krein_space_formalism:topological_classification} why this seemingly technical detail is crucial in the context of topological phenomena. 

To be able to correctly identify the nature of these symmetries, we must eliminate the superfluous degrees of freedom that were introduced when complexifying~\eqref{quantum_vs_classical:eqn:classical_real_spin_equation} while still working on a \emph{complex} Hilbert space. 

\subsubsection{Reduction to complex $\omega > 0$ states: the Schrödinger formalism for a classical spin} 
\label{quantum_vs_classical:classical:Schroedinger_formalism}
The key idea to get rid of these extraneous degrees of freedom is contained in Eq.~\eqref{quantum_vs_classical:eqn:real_solution_express_via_complex_waves}: positive and negative frequency components of the real solution $M(t) = \psi_+(t) + \psi_-(t)$ are not independent degrees of freedom as they are necessarily related by complex conjugation, $\psi_{\pm} = \overline{\psi_{\mp}}$. Therefore, 
\begin{align}
	\left (
	\begin{smallmatrix}
		a \\
		b \\
	\end{smallmatrix}
	\right ) = M = 2 \Re \psi_+ = 2 \Re \bigl ( (a - \ii b) 
	\left (
	\begin{smallmatrix}
		1 \\
		+ \ii \\
	\end{smallmatrix}
	\right )
	\bigr )
	\label{quantum_vs_classical:eqn:1_to_1_correspondence_real_complex_omega_g_0}
\end{align}
establishes a \emph{1-to-1 correspondence between real states and complex $\omega > 0$ states}, which in turn allows us to identify the \emph{real} Hilbert space $\Hil_{\R} = \R^2$ with the \emph{complex} Hilbert space $\Hil_+ = \mathrm{span} \bigl \{ 
\left (
\begin{smallmatrix}
	1 \\
	+ \ii \\
\end{smallmatrix}
\right )
\bigr \}$ composed solely of $\omega > 0$ states. Consequently, the Schrödinger form of the \emph{classical} spin Eq.~\eqref{quantum_vs_classical:eqn:classical_real_spin_equation} is obtained by restricting Eq.~\eqref{quantum_vs_classical:eqn:quantum_spin_1_2_Schroedinger_equation} to $\Hil_+$; it provides an equivalent description of the classical spin equation without introducing any superfluous, unphysical degrees of freedom. 

The implementation of the symmetries $V_j^{\R}$ on $\Hil_{\R} = \R^2$ on the complex Hilbert space $\Hil_+$ are deduced from the following two conditions: 
\begin{enumerate}[(1)]
	\item $V_j^{\R} M = 2 \Re \bigl ( V_j^{\C} \psi_+ \bigr )$ for $j = 1 , 2, 3$. 
	\item $V_j^{\C}$ is (anti)unitary on $\Hil_+$ for $j = 1 , 2 , 3$, \ie it must map $\omega > 0$ states onto $\omega > 0$ states. 
\end{enumerate}
The first condition identifies \eg $\ii \sigma_2$ and $\ii \sigma_2 C$ as candidates for $V_2^{\C}$, while the second singles out the candidate which \emph{commutes} with $H = \omega \, \sigma_2$. In this case $V_2^{\C} = \ii \sigma_2$ is realized \emph{linearly}, whereas the other two symmetries $V_{1,3}^{\C}$ are implemented as \emph{anti}unitaries. In the parlance of topological insulators $V_{1,3}^{\C}$ are \emph{even time-reversal-type symmetries} and $V_2^{\C}$ an ordinary, commuting symmetry. Our findings are summarized in Table~\ref{quantum_vs_classical:table:classical_symmetries}. 

\begin{table}
	\begin{center}
		\renewcommand{\arraystretch}{1.5}
		\newcolumntype{A}{>{\centering\arraybackslash\normalsize} m{27mm} }
		\begin{tabular}{A | A | A}
			\textbf{Real Implementation of Symmetry} & \textbf{Complex Implementation of Symmetry} & \textbf{TI Classification} \\ \hline \hline
			$V_1^{\R} = \sigma_1$     & $V_1^{\C} = \sigma_1 \, C$ & +TR \\ \hline
			$V_2^{\R} = \ii \sigma_2$ & $V_2^{\C} = \ii \sigma_2$  & ordinary \\ \hline
			$V_3^{\R} = \sigma_3$     & $V_3^{\C} = \sigma_3 \, C$ & +TR \\
		\end{tabular}
	\end{center}
	\caption{Summary of all symmetries of the quantum system. +TR stands for even time-reversal symmetry, ordinary for a commuting unitary symmetry. From the point of view of classification theory, ordinary symmetries are irrelevant. }
	\label{quantum_vs_classical:table:classical_symmetries}
\end{table}
%

\subsubsection{Physical interpretation of time-reversal-type symmetries} 
\label{quantum_vs_classical:classical:physical_interpretation_TRS}
The interpretation of $V_{1,3}^{\R,\C}$ as time-reversal-type symmetries seems to be in direct contradiction with the intuition we have of classical spins in magnetic fields: the magnetic field breaks time-reversal symmetry and the sense of rotation of the spin is determined by the sign of the magnetic field. This apparent contradiction can be resolved. 

Let us approach this from the direction of classification theory of topological insulators where the terminology stems from quantum mechanics: a time-reversal symmetry is an \emph{anti}unitary operator $T$ on a complex Hilbert space $\Hil$ that \emph{commutes} with the Hamiltonian, $T \, H = H \, T$. Therefore, it flips the arrow of time, 
\begin{align*}
	T \, \e^{- \ii t H} = \e^{- \ii (-t) H} \, T
	. 
\end{align*}
So on the basis of this definition, $V_{1,3}^{\C}$ are indeed even time-reversal-type symmetries. And as we shall see, their presence has exactly the same consequences as in solid state physics — symmetries in the band spectrum and the topological triviality of bands (\cf Section~\ref{magnonic_crystal:bulk_edge_correspondence:bulk_classification}). 

To reconcile this with our intuition, we take a closer look at the action of $V_{1,3}^{\R,\C}$: they both \emph{reverse the orientation} of the ambient space, $V_1^{\R,\C}$ exchanges $x_1$- and $x_2$-components with one another while $V_3^{\R,\C}$ corresponds to a reflection about the $x_1$-axis. So these transformations flip the sense of rotation in a coordinate system with a reversed sense of orientation — which combine to give a state that rotates in the \emph{same} direction when viewed in the original (spatial) coordinate system. \emph{From the point of view of physics the presence of these symmetries stem from the rotational symmetry of the problem.} Nevertheless, this rotational symmetry manifests itself mathematically via time-reversal-type symmetries. 

\section{The Schrödinger formalism of spin equations} 
\label{Krein_space_formalism}
The fundamental equations we are interested in are the linearization of the Landau-Lifshitz equation 
\begin{align}
	\partial_t \mathbf{M}(t) &= - \mu_0 \, \gamma \, \mathbf{M}(t) \times \Bigl ( H_0 \, e_3 + \mbox{$\sum_{j = 1}^d$} \partial_j \bigl ( A \; \partial_j \mathbf{M}(t) \bigr ) 
	+ \Bigr . \notag \\
	&\qquad \Bigl . 
	+ \mathbf{H}_{\mathrm{d}}(t) \Bigr )
	, 
	\label{Krein_space_formalism:eqn:nonlinear_Landau_Lifshitz_equations}
\end{align}
that describes the dynamics of the magnetization $\mathbf{M} = (M_1 , M_2 , M_3)$ subjected to an external magnetic field $H_0$ pointing in the $e_3$-direction. The \emph{exchange stiffness} $A(x)$ quantifies the coupling of neighboring spins, and the \emph{dipolar field} $\mathbf{H}_{\mathrm{d}}(t)$ is the solution of the magnetostatic Maxwell equations, 
\begin{subequations}\label{Krein_space_formalism:eqn:magnetostatic_Maxwell_equations}
	\begin{align}
		\nabla \times \mathbf{H}_{\mathrm{d}}(t) &= 0 
		,
		\label{Krein_space_formalism:eqn:magnetostatic_Maxwell_equations:static_dynamic}
		\\
		\nabla \cdot \bigl ( \mathbf{H}_{\mathrm{d}}(t) + \mathbf{M}(t) \bigr ) &= 0
		. 
		\label{Krein_space_formalism:eqn:magnetostatic_Maxwell_equations:constraint}
	\end{align}
\end{subequations}
Therefore, in view of the Helmholtz decomposition of vector fields the dipolar field $\mathbf{H}_{\mathrm{d}}(t) = - \nabla \Phi_{\mathrm{m}}(t)$ is a gradient field whose potential satisfies $\Delta \Phi_{\mathrm{m}}(t) = \nabla \cdot \mathbf{M}(t)$. 

In principle, we could have included other interaction terms such as the Dzyaloshinskii-Moriya interaction \cite{Mook:magnon_hall_effect_kagome:2014,Mook:topological_magnon_edge_state:2014,Mook:topological_magnon_kagome_interface:2015, owerre:magnon_haldane:2016,kim_et_al:magnon_haldane_kane_melel:2016}, but for the sake of brevity and concreteness, we chose to consider the same equations as Shindou et al.\ \cite{Shindou_et_al:chiral_magnonic_edge_modes:2013}. However, we emphasize that our framework is much more general and \eg applies to ferromagnetic and antiferromagnetic systems alike. 

Now let us linearize the Landau-Lifshitz equations around the saturation magnetization 
\begin{align*}
	\mathbf{M} = M_{\mathrm{s}} \, e_3 + \mathbf{m}
	, 
\end{align*}
where $\mathbf{m} = (m_1 , m_2 , m_3)$ is the small deviation from $M_{\mathrm{s}} \, e_3$ that we would like to study. Because the linearized equation conserves $M_3 \approx M_{\mathrm{s}}$, we drop the $e_3$-component from the equations and arrive at 
\begin{align}
	\ii \partial_t m(t) &= H m(t) 
	\label{Krein_space_formalism:eqn:real_Schroedinger_Krein_form_linearized_LL_equations}
\end{align}
where $m = (m_1 , m_2)$ are the first two components of the spin and the “hamiltonian”
\begin{align}
	H = W^{-1} \, (D + L)
	\label{Krein_space_formalism:eqn:spin_hamiltonian}
\end{align}
is the product of the “weight operator”
\begin{align}
	W^{-1} = \sigma_2 \, M_{\mathrm{s}}
\end{align}
with the sum of the two operators 
\begin{subequations}\label{Krein_space_formalism:eqn:contributions_spin_hamiltonian}
	\begin{align}
		D &= \mu_0 \, \gamma \, M_{\mathrm{s}}^{-1} \, \bigl ( H_0 + \nabla A \cdot \nabla M_{\mathrm{s}} + A \; \Delta M_{\mathrm{s}} \bigr ) 
		\, + \notag \\
		&\qquad 
		- \mu_0 \, \gamma \, \bigl ( \nabla A \cdot \nabla + A \, \Delta \bigr )
		, 
		\label{Krein_space_formalism:eqn:contributions_spin_hamiltonian:K}
		\\
		L &= - \ii \, \mu_0 \, \gamma \, \sigma_2 \, \biggl ( \sigma_3 \, \partial_1 \partial_2
		\, + \biggr . 
		\notag 
		\\
		&\qquad \biggl . 
		+ \frac{\sigma_1 + \ii \sigma_2}{2} \, \partial_2^2 - \frac{\sigma_1 - \ii \sigma_2}{2} \, \partial_1^2 \biggr ) \; \Delta^{-1}
		. 
		\label{Krein_space_formalism:eqn:contributions_spin_hamiltonian:L}
	\end{align}
\end{subequations}
The reason why we split $H$ in this particular way will become clear in the next subsection. Later in Section~\ref{magnonic_crystals} we will assume that $A(x)$, $H_0(x)$ and $M_{\mathrm{s}}(x)$ are periodic, but this is not necessary at this point. 
\medskip

\noindent
Rewriting the linearized spin equations~\eqref{Krein_space_formalism:eqn:real_Schroedinger_Krein_form_linearized_LL_equations} in the mathematical formalism of a quantum Schrödinger equation goes way beyond expressing Eq.~\eqref{Krein_space_formalism:eqn:real_Schroedinger_Krein_form_linearized_LL_equations} in terms of the “hamiltonian”~\eqref{Krein_space_formalism:eqn:spin_hamiltonian}. The crucial piece here is to (1)~identify the vector space $\Hil_{\R}$ of real-valued spins $m = (m_1 , m_2)$ in terms of a complex vector space $\Hil_{\C}$; and (2)~find a scalar product on $\Hil_{\C}$ so that the spin “hamiltonian” $H$ is hermitian on $\Hil_{\C}$. 
\begin{figure*}
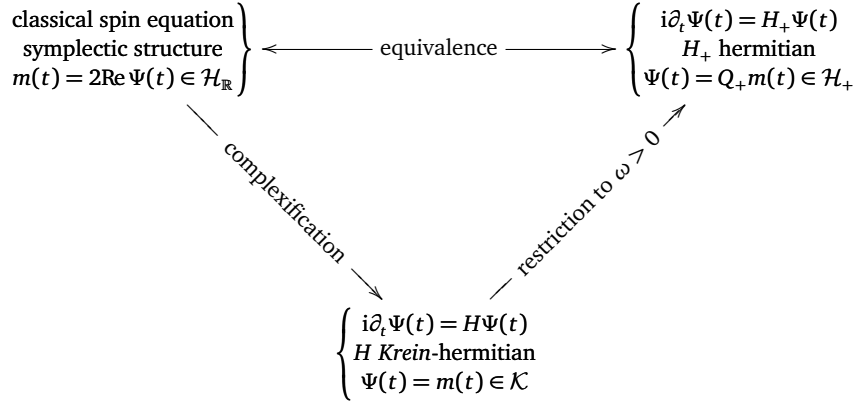

	\begin{align*}
		\bfig
			\node classical(-2400,0)[
			\left .
			\begin{matrix}
				\mbox{classical spin equation} \\
				\mbox{symplectic structure} \\
				m(t) = 2 \Re \Psi(t) \in \Hil_{\R} \\
			\end{matrix}
			\right \}
			]
			\node complexification(-1200,-1200)[
			\left \{
			\begin{matrix}
				\ii \partial_t \Psi(t) = H \Psi(t) \\
				\mbox{$H$ \emph{Krein}-hermitian} \\
				\Psi(t) = m(t) \in \Krein \\
			\end{matrix}
			\right .
			]
			\node Schroedinger(0,0)[
			\left \{
			\begin{matrix}
				\ii \partial_t \Psi(t) = H_+ \Psi(t) \\
				\mbox{$H_+$ hermitian} \\
				\Psi(t) = Q_+ m(t) \in \Hil_+ \\
			\end{matrix}
			\right .
			]
			\arrow|m|/<->/[classical`Schroedinger;\mbox{\small ~equivalence~}]
			\arrow|m|[classical`complexification;\rotatebox{-45}{\mbox{\small complexification}}]
			\arrow|m|[complexification`Schroedinger;\rotatebox{45}{\mbox{\small restriction to $\omega > 0$}}]
		\efig
	\end{align*}
	\caption{The goal is to establish the one-to-one correspondence between the original equations defined for real states on the top-left and the representation as complex spin waves on the top-right. The symplectic structure \cite[Chapter~8.1]{Goldstein:classical_mechanics:2013} of the real spin equation gives rise to a weighted scalar product on the top-right. The derivation is done in two \emph{unidirectional} steps where we first complexify the real equations and then discard all negative frequency states. }
	\label{Krein_space_formalism:fig:strategy_diagram}
\end{figure*}

The strategy we implement below is summarized in Figure~\ref{Krein_space_formalism:fig:strategy_diagram}: corresponding to the arrow from the top-left to the bottom, we first complexify the classical equations to be able to use complex exponentials, which emerge naturally in the context of first-order equations. This effectively doubles the degrees of freedom and adds unphysical states that have a non-zero imaginary part. This complex vector space can be furnished with an indeterminate \emph{Krein} inner product and becomes a \emph{Krein space}; here, the length of a non-zero vector can be positive, negative or zero. Associated with this Krein inner product is a \emph{Krein adjoint}; with respect to this Krein adjoint, the spin “hamiltonian” is Krein hermitian. 

Provided the spin equations are thermodynamically stable (\ie the spin “hamiltonian” satisfies Eq.~\eqref{Krein_space_formalism:eqn:stability_condition}), the second step eliminates all the superfluous degrees of freedom by representing the real spin vector $m = 2 \Re \Psi$ as the real part of a complex positive frequency wave. For positive frequency waves, the Krein inner product is positive definite again and therefore defines a scalar product — the positive frequency subspace of the Krein space is a Hilbert space. 

While these two steps are one way, it turns out that their concatenation is reversible as indicated by the left-right arrow going from the top-left to the top-right in Figure~\ref{Krein_space_formalism:fig:strategy_diagram}.

\subsection{Krein space structure of the complexified equations and reduction to a Schrödinger-type equation} 
\label{Krein_space_formalism:structure}
The first step towards obtaining the Schrödinger form of Eq.~\eqref{Krein_space_formalism:eqn:real_Schroedinger_Krein_form_linearized_LL_equations} is to admit complex spin density vectors $\Psi$, which are assumed to be elements of the complex Hilbert space 
\begin{align*}
	\Hil(\R^d,\C^2) &= \Bigl \{ \Psi \; \; \big \vert \; \; \int_{\R^d} \dd x \, \abs{\Psi(x)}^2 < \infty \Bigr \} 
	, 
\end{align*}
where $d$ is the dimension of the ambient space, typically $d = 1 , 2 , 3$. Physically speaking, this Hilbert space is comprised of localized excitations with respect to $M_{\mathrm{s}} \, e_3$. The scalar product is the usual one, 
\begin{align*}
	\scpro{\Phi}{\Psi} &= \int_{\R^d} \dd x \, \Phi(x) \cdot \Psi(x) 
	, 
\end{align*}
where complex conjugation is contained in the dot product $\Phi(x) \cdot \Psi(x) = \overline{\Phi_1(x)} \, \Psi_1(x) + \overline{\Phi_2(x)} \, \Psi_2(x)$. 

A major mathematical difference between the simple spin hamiltonian $H = \omega \, \sigma_2$ covered in Section~\ref{quantum_vs_classical} and $H \neq H^{\dagger}$ given by \eqref{Krein_space_formalism:eqn:spin_hamiltonian} is that the latter is \emph{not} a hermitian operator. Instead, it is a Krein- or para-hermitian operator \footnote{Throughout this paper, we shall use the moniker “Krein-” rather than “para-”, which is more in line with the mathematics literature.}.

\subsubsection{A primer on Krein spaces} 
\label{Krein_space_formalism:structure:primer}
For the benefit of the reader, we will first recap the basics of the theory of Krein spaces. A Krein space $\Krein$ is a Hilbert space $\mathcal{H}$ equipped with a second, \emph{Krein inner product}
\begin{align}
	\bscpro{\Phi}{\Psi}_W = \bscpro{\Phi}{W \, \Psi}
	, 
	\label{Krein_space_formalism:eqn:Krein_inner_product}
\end{align}
that has all the properties of a scalar product — conjugate symmetry and linearity in the second entry — except for positive definiteness \cite[Definition~1.1.2]{Tomita:intro_Krein_spaces:1980}. Here, the weight $W = W^{\dagger}$ is a hermitian, bounded operator with bounded inverse, and $\sscpro{\, \cdot \,}{\, \cdot \,}$ is the scalar product that came with the Hilbert space $\Hil$. Choosing $W = 
\left (
\begin{smallmatrix}
	1 & 0 \\
	0 & -\id \\
\end{smallmatrix}
\right )$ on $\C^4$, for example, gives rise to the Lorentzian metric $\norm{\Psi}_W^2 = \scpro{\Psi}{\Psi}_W = \sabs{\psi_0}^2 - \sum_{j = 1}^3 \sabs{\psi_j}^2$. Thus, if we use the Krein inner product to measure distances, the length of a non-zero vector may be positive, negative or even $0$. 

This Krein inner product gives rise to a notion of adjoint 
\begin{align*}
	A^{\sharp} = W^{-1} \, A^{\dagger} \, W 
	, 
\end{align*}
which by definition satisfies $\bscpro{A^{\sharp} \Phi}{\Psi}_W = \bscpro{\Phi}{A \Psi}_W$. A Krein-hermitian operator $H^{\sharp} = H$ is one that is hermitian with respect to the Krein inner product \footnote{To avoid having to deal with mathematical minutiæ, we will not distinguish between (Krein-)hermitian and (Krein-)selfadjoint operators.}, and a Krein-unitary $U$ is an operator whose Krein-adjoint 
\begin{align*}
	U^{\sharp} = W^{-1} \, U^{\dagger} \, W \overset{!}{=} U^{-1} 
\end{align*}
coincides with its inverse. Krein-\emph{anti}unitary operators are invertible, antilinear operators that satisfy 
\begin{align*}
	\bscpro{U \Phi}{U \Psi}_W = \overline{\scpro{\Phi}{\Psi}_W} = \scpro{\Psi}{\Phi}_W 
	. 
\end{align*}
The concepts of Krein-adjoint and Krein-unitary generalize to maps between two different Krein spaces in a straightforward fashion. 

Similarly, a Krein-orthogonal projection is an operator $P^2 = P = P^{\sharp}$ that squares to itself and is Krein-hermitian; the latter condition $P = W^{-1} \, P^{\dagger} \, W$ states that $P$ and $P^{\dagger}$ are related by a similarity transform. 

Note that Krein-hermiticity places much weaker conditions on the operator than hermiticity. For instance, the spectrum of Krein-hermitian operators, \ie the set of complex numbers $z$ such that $H - z$ is not invertible, need not be real. Indeed, $H = H^{\sharp}$ may have complex eigenvalues and if in addition the spectrum consists only of eigenvalues, the associated eigenvectors need not form a basis. 

However, a special class of operators, so-called \emph{Krein-spectral operators} \cite[Definition~3.2.1]{Tomita:intro_Krein_spaces:1980}, will still retain all features of hermitian operators: these are Krein-hermitian operators $H = H^{\sharp}$ which are Krein-unitarily equivalent to an operator 
\begin{align*}
	\widetilde{H} = U \, H \, U^{-1} 
	= U \, H \, U^{\sharp} 
	= \widetilde{H}^{\sharp}
	= \widetilde{H}^{\dagger}
\end{align*}
that is hermitian in the ordinary sense \emph{and} Krein-hermitian. Operators which are positive with respect to the Krein inner product, \ie those that satisfy $\bscpro{\Psi}{H \Psi}_W > 0$ whenever $\Psi \neq 0$, are automatically Krein-spectral (\cf \cite[Proposition~2]{Peano_Schulz_Baldes:topological_edge_states_bosonic_systems:2018} and \cite{Colpa:diagonalization_boson_hamiltonian:1978}). 

With these notions in hand, let us get back to spin Eq.~\eqref{Krein_space_formalism:eqn:real_Schroedinger_Krein_form_linearized_LL_equations}. 

\subsubsection{Hamiltonian~\eqref{Krein_space_formalism:eqn:spin_hamiltonian} as a Krein-hermitian operator} 
\label{Krein_space_formalism:structure:spin_Krein_hermitian}
The purpose of this primer was to introduce the right mathematical framework in which to treat Eq.~\eqref{Krein_space_formalism:eqn:real_Schroedinger_Krein_form_linearized_LL_equations}. Let us introduce the indeterminate inner product 
\begin{align}
	\sscpro{\Phi}{\Psi}_W &= \bscpro{\Phi \,}{\, M_{\mathrm{s}}^{-1} \, \sigma_2 \, \Psi}
	\label{Krein_space_formalism:eqn:concrete_Krein_inner_product}
	\\
	&= \int_{\R^d} \dd x \, \Phi(x) \cdot M_{\mathrm{s}}^{-1}(x) \, \sigma_2 \Psi(x) 
	\notag 
\end{align}
on the Hilbert space $\Hil(\R^d,\C^2)$; since the saturation magnetization $M_{\mathrm{s}}$ is assumed to be non-zero everywhere, $\sabs{M_{\mathrm{s}}(x)} \geq M_0 > 0$, the bilinear form~\eqref{Krein_space_formalism:eqn:concrete_Krein_inner_product} is indeed a Krein inner product. We shall denote the resulting Krein space $\bigl ( \Hil(\R^d,\C^2) \, , \, \sscpro{\, \cdot \,}{\, \cdot \,}_W \bigr )$ with $\Krein(\R^d,\C^2)$. 

A straightforward, but somewhat lengthy computation shows that with respect to the Krein inner product~\eqref{Krein_space_formalism:eqn:concrete_Krein_inner_product} the operator $H$ from \eqref{Krein_space_formalism:eqn:spin_hamiltonian} is indeed Krein-hermitian: first of all, the Krein-hermiticity of $H = W^{-1} \, (D + L)$ is equivalent to the hermiticity of $D$ and $L$. 

The hermiticity of the first terms 
\begin{align*}
	\mu_0 \, \gamma \, M_{\mathrm{s}}^{-1} \, &\bigl ( H_0 + \nabla A \cdot \nabla M_{\mathrm{s}} + A \; \Delta M_{\mathrm{s}} \bigr ) 
	= \\
	&= \Bigl ( \mu_0 \, \gamma \, M_{\mathrm{s}}^{-1} \, \bigl ( H_0 + \nabla A \cdot \nabla M_{\mathrm{s}} + A \; \Delta M_{\mathrm{s}} \bigr ) \Bigr )^{\dagger} 
\end{align*}
is immediate as the derivatives do not act on the spin waves; for the second, partial integration gives us 
\begin{align*}	
	\bigl ( \nabla A \cdot \nabla + A \; \Delta \bigr ) &= \bigl ( \nabla A \cdot \nabla + A \; \Delta \bigr )^{\dagger} 
	. 
\end{align*}
To verify $L^{\dagger} = L$, we exploit that $- \ii \, \sigma_2 = - \bigl ( - \ii \, \sigma_2 \bigr )^{\dagger}$ \emph{anti}commutes with with the term in parentheses of \eqref{Krein_space_formalism:eqn:contributions_spin_hamiltonian:L} up to an exchange of the roles of $\partial_1$ and $\partial_2$. Furthermore, this second factor of $L$ is hermitian up to a \emph{second} exchange of $\partial_1$ and $\partial_2$. Putting both pieces together yields $(W^{-1} \, L)^{\sharp} = W^{-1} \, L$. 

When the system is thermodynamically stable, \ie when $\bscpro{\Psi}{H \Psi}_W > 0$ whenever $\Psi \neq 0$, then $H$ is Krein-spectral and can therefore be block-diagonalized via a Krein unitary \cite{Colpa:diagonalization_boson_hamiltonian:1978}. This is exactly the situation studied in \cite{Colpa:diagonalization_boson_hamiltonian:1978,Shindou_et_al:chiral_magnonic_edge_modes:2013,Peano_Schulz_Baldes:topological_edge_states_bosonic_systems:2018}, which yields real frequency bands and a complete basis of Bloch functions when $M_{\mathrm{s}}(x)$ and $A(x)$ are periodic. 

\subsubsection{Comparison to Shindou et al.\ \cite{Shindou_et_al:chiral_magnonic_edge_modes:2013}} 
\label{Krein_space_formalism:comparison}
While works such as \cite{Shindou_et_al:chiral_magnonic_edge_modes:2013} already \emph{implicitly} make use of the Krein space formalism, they do not exploit the advantages the Krein structure affords to the fullest extent. 

First of all, Shindou et al.\ use a different representation than we do: they work in the eigenbasis of $\sigma_2$ and absorb a factor of $\nicefrac{1}{\sqrt{M_{\mathrm{s}}}}$ into the definition of the spin waves (\cf \cite[Eq.~(18)]{Shindou_et_al:chiral_magnonic_edge_modes:2013}). This gets rid of the factor $M_{\mathrm{s}}^{-1}$ in the inner product and replaces $\sigma_2$ with $\sigma_3$. 

Moreover, they write out the Krein adjoint. For example, the operator $T_{\mathbf{k}}$ from \cite[Eq.~(3)]{Shindou_et_al:chiral_magnonic_edge_modes:2013} is a Krein-unitary, and $H_{\mathbf{k}}$ from \cite[Eq.~(2)]{Shindou_et_al:chiral_magnonic_edge_modes:2013} is a Krein-spectral operator as the right-hand side can be written as $\sigma_3 \, 
\left (
\begin{smallmatrix}
	E_{\mathbf{k}} & 0 \\
	0 & E_{\mathbf{-k}} \\
\end{smallmatrix}
\right )$. The existence of a diagonalizing Krein unitary follows from the \emph{thermodynamic stability} assumption. 

However, there is a subtle difference between the definition of the hamiltonian: the operator $H_{\mathbf{k}}$ used by Shindou et al.\ is not Krein hermitian as they factor out $\sigma_3$ (\cf \cite[Eq.~(21)]{Shindou_et_al:chiral_magnonic_edge_modes:2013}). While ultimately this is merely a matter of convention — we prefer to reserve the letter $H$ for the generator of the dynamics via \eqref{Krein_space_formalism:eqn:real_Schroedinger_Krein_form_linearized_LL_equations}, Shindou et al.\ instead denote with $H$ the single-particle hamiltonian that enters into the definition of the quadratic boson hamiltonian \cite[Eq.~(1)]{Shindou_et_al:chiral_magnonic_edge_modes:2013}. 

Similarly, thanks to the Krein formalism it is immediately clear why the operator 
\begin{align*}
	\mathbf{P}_j &= \mathbf{T}_{\mathbf{k}} \, \pmb{\Gamma}_j \, \sigma_3 \, \mathbf{T}_{\mathbf{k}}^{\dagger} \, \sigma_3
	= \mathbf{T}_{\mathbf{k}} \, \pmb{\Gamma}_j \, \mathbf{T}_{\mathbf{k}}^{\sharp} 
	= \mathbf{T}_{\mathbf{k}} \, \pmb{\Gamma}_j \, \mathbf{T}_{\mathbf{k}}^{\, -1} 
	, 
\end{align*}
as defined in \cite[Eq.~(6)]{Shindou_et_al:chiral_magnonic_edge_modes:2013} is \emph{the} Krein-orthogonal projection onto the $j$th frequency band. 

\subsection{Symmetries of the complexified equations} 
\label{Krein_space_formalism:symmetries}
The operator $H$ comes furnished with a particle-hole-type \emph{constraint}, $\{ H , C \} = 0$, as it arises from Eq.~\eqref{Krein_space_formalism:eqn:real_Schroedinger_Krein_form_linearized_LL_equations} which describes real-valued waves. As explained in Section~\ref{quantum_vs_classical:classical:complexification} complex conjugation $C$ enters as a \emph{constraint} and does not correspond to a symmetry of the physical system. That is because all physically relevant states are real, and complex conjugation acts trivially on physical, real-valued spin waves $m(t) = C m(t)$. 

Apart from complex conjugation and crystallographic symmetries, there are six other candidates for physical symmetries, the three Pauli matrices $\sigma_1$, $\ii \sigma_2$ and $\sigma_3$ as well as $\sigma_1 \, C$, $\ii \sigma_2 \, C$ and $\sigma_3 \, C$. On the level of the complexified equations $\sigma_j$ and $\sigma_j \, C$ \emph{implement the same physical symmetry}. So to choose amongst these (anti)unitaries, let us pick those that are also \emph{Krein}-(anti)unitaries as well, namely 
\begin{align*}
	V_{1,3}^{\C} &= \sigma_{1,3} \, C 
	, 
	\\
	V_2^{\C} &= \ii \sigma_2 
	, 
\end{align*}
which commute with the weights $W = M_{\mathrm{s}}^{-1} \, \sigma_2$. 

As before, $V_{1,3}^{\C}$ are even time-reversal-type symmetries that are also \emph{Krein}-antiunitaries; $V_2^{\C}$ is an ordinary symmetry that is also a Krein-unitary. 

Consequently, all three of the $V_j^{\C}$'s commute with $W^{-1} \, D$, the first term that makes up $H = W^{-1} \, (D + L)$. The second one, though, breaks all three symmetries: if we keep track of the order of the derivatives and use $L(\partial_1,\partial_2)$ for operator~\eqref{Krein_space_formalism:eqn:contributions_spin_hamiltonian:L}, then the action of the symmetry operators computes to 
\begin{align*}
	V_1^{\C} \, L(\partial_1 , \partial_2) \, V_1^{\C} &= + L(\partial_2 , \partial_1)
	, 
	\\
	V_2^{\C} \, L(\partial_1 , \partial_2) \, V_2^{\C} &= - L(\partial_2 , \partial_1)
	, 
	\\
	V_3^{\C} \, L(\partial_1 , \partial_2) \, V_3^{\C} &= + L(\partial_1 , -\partial_2)
	= + L(-\partial_1 , \partial_2)
	. 
\end{align*}
Thus, $L$ breaks all three symmetries and $H$ possesses no time-reversal-type symmetries. The lack of a time-reversal symmetry is a \emph{necessary} condition for there to be non-trivial topological effects captured by Chern numbers. Moreover, we once again see that the absence of time-reversal-type symmetries are related to the lack of rotational symmetry of the physical system. 

Other symmetry actions such as translations and rotations can be treated in the same fashion: first consider the symmetry actions as operators on the Krein space $\Krein(\R^d,\C^2)$, with and without complex conjugation. Amongst the two, we choose the symmetry action which is simultaneously an (anti)unitary \emph{and} a \emph{Krein}-(anti)unitary. For lattice translations we shall do that in Section~\ref{magnonic_crystals:changing_representation:fourier_Krein}. 

\subsection{The Schrödinger formalism of the classical spin equation~\eqref{Krein_space_formalism:eqn:real_Schroedinger_Krein_form_linearized_LL_equations}} 
\label{Krein_space_formalism:Schroedinger_formalism}
While what we did up to now is pretty standard in the mathematical physics literature (see \eg \cite{Schulz_Baldes_Villegas_Blas:signatures_operators_real_Krein_spaces:2017,Peano_Schulz_Baldes:topological_edge_states_bosonic_systems:2018}), what we add here to the discussion is to properly include the real-valuedness of the spin waves. And indeed, the Krein structure will \emph{not} play a role at the end for systems that are \emph{thermodynamically stable}, 
\begin{align}
	\sscpro{\Psi}{H \, \Psi}_W = \bscpro{\Psi \, }{(D + L) \, \Psi} > 0 
	, 
	\label{Krein_space_formalism:eqn:stability_condition}
\end{align}
where the “expectation value” of $H$ with respect to the \emph{Krein inner product} is strictly positive for all $\Psi \neq 0$. \emph{We shall assume from hereon that $H$ is thermodynamically stable.} Then $H$ is Krein-unitarily equivalent to a \emph{hermitian} operator, which implies among other things that the spectrum of $H$ is real. 

Note that all of our arguments also transfer to the case when we study the second quantization of \eqref{Krein_space_formalism:eqn:real_Schroedinger_Krein_form_linearized_LL_equations} as is commonly done in the literature, because creation and annihilation operators are of course not “independent variables” (just like $\Psi_+$ and $\Psi_- = C \Psi_+$). We will address this point in depth in Section~\ref{second_quantized_equations:particle_hole_constraint}.

\subsubsection{Reduction to complex $\omega > 0$ spin waves} 
\label{Krein_space_formalism:Schroedinger_formalism:reduction_to_complex_omega_0_spin_waves}
For thermodynamically stable systems, we can exploit the reality of 
\begin{align}
	m(t) = C m(t) = \Psi_+(t) + \Psi_-(t) = 2 \Re \Psi_{\pm}(t) 
	\label{Krein_space_formalism:eqn:connection_real_complex_wave}
\end{align}
just like for the single classical spin we studied in Section~\ref{quantum_vs_classical:classical}, and write the real-valued spin wave as the real part of a complex $\omega > 0$ wave. That is because the negative frequency contribution $\Psi_-(t) = C \Psi_+(t)$ is redundant. In fact, we once again have a \emph{1-to-1 correspondence}
\begin{align}
	\bfig
		\node classical(0,0)[\Hil_{\R}(\R^d,\C^2) \ni m(t) = 2 \Re \Psi_{\pm}(t)]
		\node quantum_rep(0,-350)[\Psi_{\pm}(t) = Q_{\pm} m(t) \in \Krein_{\pm}(\R^d,\C^2)]
		\arrow|m|/<->/[classical`quantum_rep;]
	\efig
	\label{Krein_space_formalism:eqn:1_to_1_correspondence}
\end{align}
between elements of the \emph{real subspace}
\begin{align*}
	\Hil_{\R}(\R^d,\C^2) &= 2 \Re \, \Krein 
	= \Hil(\R^d,\R^2) 
	\\
	&= \Bigl \{ \mbox{$\Psi$ real-valued} \; \; \big \vert \; \; \int_{\R^d} \dd x \, \sabs{\Psi(x)}^2 < \infty \Bigr \} 
\end{align*}
and the \emph{positive/negative frequency subspace}
\begin{align*}
	\Krein_{\pm}(\R^d,\C^2) &= Q_{\pm} \, \Krein(\R^d,\C^2)
	\\
	&= \Bigl \{ \mbox{$\Psi$ is $\pm \omega > 0$ wave} \; \; \big \vert \; \; \int_{\R^d} \dd x \, \sabs{\Psi(x)}^2 < \infty \Bigr \} 
	. 
\end{align*}
Here, the Krein-orthogonal projection onto the positive/negative frequencies 
\begin{align}
	Q_{\pm} = Q_{\pm}^{\sharp}
	= \frac{\ii}{2 \pi} \int_{\Gamma_{\pm}} \dd z \, (H - z)^{-1} 
	\label{Krein_space_formalism:eqn:positive_frequency_Krein_projection}
\end{align}
can be defined through a contour integral where $\Gamma_{\pm}$ encloses the positive/negative part of the spectrum of $H$. When $H$ is periodic, we can alternatively express $Q_{\pm}$ in terms of positive/negative frequency Bloch functions. 

By convention, we choose to represent $m = 2 \Re \Psi_+$ in terms of $\omega > 0$ waves. 

One succinct way to express Eq.~\eqref{Krein_space_formalism:eqn:1_to_1_correspondence} — which underlies the top-most equivalence in Figure~\ref{Krein_space_formalism:fig:strategy_diagram} — is to say that $Q_{\pm}$ and $2 \Re$ are inverses of one another, 
\begin{align*}
	2 \Re \, Q_{\pm} \, \Re &= \Re 
	,
	\\
	Q_{\pm} \, 2 \Re &= \id_{\Krein_{\pm}(\R^d,\C^2)}
	. 
\end{align*}
These equalities follow from $C \, Q_{\pm} \, C = Q_{\mp}$ and $Q_+ + Q_- = \id$. Each of the above equalities represents one direction of the arrow: the first one starting from the upper-left of Figure~\ref{Krein_space_formalism:fig:strategy_diagram} going to the upper-right and returning to the upper-left, and vice versa for the second one. 

\subsubsection{The positive frequency space $\Krein_+(\R^d,\C^2) = \Hil_+(\R^d,\C^2)$ as a Hilbert space with weighted scalar product} 
\label{Krein_space_formalism:Schroedinger_formalism:K_plus_Hilbert_space}
$\Krein_+(\R^d,\C^2)$ in principle inherits the Krein structure from its parent $\Krein(\R^d,\C^2)$, and we may consider $\scpro{\, \cdot \,}{\, \cdot \,}_W$ only on $\Krein_+(\R^d,\C^2)$. However, it turns out \cite[Proposition~2]{Peano_Schulz_Baldes:topological_edge_states_bosonic_systems:2018} that on $\Krein_+(\R^d,\C^2)$ the Krein inner product 
\begin{align*}
	\scpro{\Psi_+}{\Psi_+}_W \geq 0 
\end{align*}
has a fixed sign, and equals $0$ if and only if $\Psi_+ = 0$. Therefore, the Krein inner product $\scpro{\, \cdot \,}{\, \cdot \,}_W$ is really just an alternative \emph{scalar} product in addition to the usual, unweighted one $\scpro{\, \cdot \,}{\, \cdot \,}$. Thus, we may forget about the unweighted scalar product and regard $\Krein_+(\R^d,\C^2)$ as a \emph{Hilbert} space; we will denote this Hilbert space with $\Hil_+(\R^d,\C^2)$. 

Consequently, the restriction 
\begin{align*}
	H_+ = H \vert_{\omega > 0} = Q_+ \, H \, Q_+ 
\end{align*}
of the “hamiltonian” to $\Hil_+(\R^d,\C^2)$ is \emph{hermitian} with respect to $\scpro{\, \cdot \,}{\, \cdot \,}_W$ and thanks to the 1-to-1 correspondence~\eqref{Krein_space_formalism:eqn:1_to_1_correspondence} of real and complex $\omega > 0$ waves we may equivalently formulate \eqref{Krein_space_formalism:eqn:real_Schroedinger_Krein_form_linearized_LL_equations} as a \emph{Schrödinger equation in the mathematical sense}, 
\begin{align}
	\ii \partial_t \Psi_+(t) &= H_+ \Psi_+(t) 
	, 
	\label{Krein_space_formalism:eqn:Schroedinger_form_spin_wave_equation}
	\\
	\Psi_+(t_0) &= Q_+ m(t_0) \in \Hil_+(\R^d,\C^2)
	. 
	\notag 
\end{align}
By that we mean that the generator of the dynamics $H^{\sharp} = H$ is a hermitian operator acting on a \emph{Hilbert} space $\Hil_+(\R^d,\C^2)$. We do \emph{not} mean to imply that the physical interpretation from quantum mechanics carries over, so $\Psi_+(t)$ still just describes a \emph{classical} wave, which in principle can be directly observed in experiment. Nevertheless, because we treat the classical spin equation in the mathematical framework of quantum mechanics, we can systematically employ techniques initially developed for quantum systems to spin waves. 

\emph{All of the physics of spin waves is contained in Eq.~\eqref{Krein_space_formalism:eqn:Schroedinger_form_spin_wave_equation} since it is equivalent to the linearized spin equation~\eqref{Krein_space_formalism:eqn:real_Schroedinger_Krein_form_linearized_LL_equations}.} So if $m(t)$ solves the classical spin equation~\eqref{Krein_space_formalism:eqn:real_Schroedinger_Krein_form_linearized_LL_equations}, then $\Psi_+(t) = Q_+ m(t)$ is a solution to Eq.~\eqref{Krein_space_formalism:eqn:Schroedinger_form_spin_wave_equation}. Conversely, solutions $\Psi_+(t)$ to Eq.~\eqref{Krein_space_formalism:eqn:Schroedinger_form_spin_wave_equation} give rise to solutions $m(t) = 2 \Re \Psi_+(t)$ of Eq.~\eqref{Krein_space_formalism:eqn:real_Schroedinger_Krein_form_linearized_LL_equations}. 

\subsubsection{Implementing symmetries on $\Hil_+(\R^d,\C^2)$} 
\label{Krein_space_formalism:Schroedinger_formalism:symmetries}
When we were given the choice to decide between implementing one of the three symmetries discussed in Section~\ref{Krein_space_formalism:symmetries}, we opted for the realization which is also a \emph{Krein}-(anti)unitary on the level of operators. We will now explain why the \emph{Krein}-(anti)unitary implementation is more suitable: it restricts to an (anti)unitary symmetry on $\Hil_+(\R^d,\C^2)$. 

Suppose $U$ is a Krein-(anti)unitary on the Krein space $\Krein(\R^d,\C^2)$ which \emph{commutes} with $H$. Then in view of Eq.~\eqref{Krein_space_formalism:eqn:positive_frequency_Krein_projection} the operator $U$ commutes with $Q_+$ and therefore maps $\omega > 0$ states onto $\omega > 0$ states. Moreover, its restriction $U_+ = U \vert_{\omega > 0}$ to $\Hil_+(\R^d,\C^2)$ inherits $U_+^{-1} = U_+^{\sharp}$ and is therefore a \emph{unitary} on $\Hil_+(\R^d,\C^2)$ with respect to the \emph{scalar} product $\scpro{\, \cdot \,}{\, \cdot \,}_W$. 

If $U$ is an (anti)unitary, but not a \emph{Krein}-(anti)unitary, then it does not correspond to a physical symmetry: since the classical wave equation can be written as the Schrödinger-type Eq.~\eqref{Krein_space_formalism:eqn:Schroedinger_form_spin_wave_equation} with scalar product $\scpro{\, \cdot \,}{\, \cdot \,}_W$ on $\Hil_+(\R^d,\C^2)$, by Wigner's theorem \cite{Wigner:group_theory_quantum_mechanics:1959,Bracci_Morchio_Strocchi:Wigners_theorem_Krein_spaces:1975} symmetries \emph{have to be} implemented unitarily or antiunitarily with respect to $\scpro{\, \cdot \,}{\, \cdot \,}_W$. 

(Anti)unitaries $U$ which \emph{anti}commute with $H$ are not compatible with the restriction to $\omega > 0$ states, because such operators intertwine $Q_+$ with the analogously defined Krein-orthonogal projection $Q_- = U \, Q_+ \, U^{\sharp}$ onto the negative frequency states. However, in that case $U \, C$ then commutes with $H$ and is therefore a \emph{candidate} for a symmetry. If $U \, C$ is also a Krein-(anti)unitary, then it implements a symmetry of the classical spin equations. 

\subsection{Topological classification of $H_+ = H \vert_{\omega > 0}$ as class~A} 
\label{Krein_space_formalism:topological_classification}
The payoff for using the Schrödinger formalism of classical spin waves is that we can apply the Cartan-Altland-Zirnbauer classification scheme of topological insulators \cite{Altland_Zirnbauer:superconductors_symmetries:1997,Chiu_Teo_Schnyder_Ryu:classification_topological_insulators:2016} to spin waves. 

The specific spin wave “hamiltonian” considered here, Eq.~\eqref{Krein_space_formalism:eqn:spin_hamiltonian}, lacks time-reversal-type and anticommuting symmetries. Consequently, in the parlance of \cite{Altland_Zirnbauer:superconductors_symmetries:1997,Chiu_Teo_Schnyder_Ryu:classification_topological_insulators:2016} $H_+$ is an \emph{operator of class~A}, the same topological class that operators describing the (Integer) Quantum Hall Effect belong to. So in this precise sense, \emph{Shindou et al.\ really did propose in \cite{Shindou_et_al:chiral_magnonic_edge_modes:2013} an analog of the Quantum Hall Effect for spin waves in magnonic crystals.} 

The step of eliminating superfluous degrees of freedom, or, equivalently, distinguishing between $C$ as a (particle-hole-type) symmetry and a constraint, was crucial for otherwise we would have surmised that the spin Eq.~\eqref{Krein_space_formalism:eqn:real_Schroedinger_Krein_form_linearized_LL_equations} is of \emph{class~D} rather than \emph{class~A} as the BdG “hamiltonian” $H$ possesses a particle-hole-type “symmetry”. Because class~D operators possess additional topological invariants that are not supported in class~A, this distinction is \emph{not purely academic} and \emph{manifests itself physically.} 

Moreover, and perhaps from a theoretical point of view \emph{unfortunately}, the Krein structure does not play a role for the topological classification. Otherwise, there might have been novel topological phenomena due to commuting and anticommuting Krein-(anti)unitaries that give rise to a more complicated zoology of topological insulators \cite{DeNittis_Gomi:K_theoretic_classification_operators_on_Krein_spaces:2019}. 

However, if all we are interested in the spin wave analog of the Quantum Hall Effect, our results are good news: $H_+$ is of class~A, and we expect that the same bulk-edge correspondence quantitatively predicts the net number of magnonic edge modes. And the absence of time-reversal-type symmetries in $H$ and $H_+$ defined through Eq.~\eqref{Krein_space_formalism:eqn:spin_hamiltonian} is a necessary but not sufficient condition for the existence of topologically protected edge modes. 

We could have included other interactions such as Dzyaloshinskii-Moriya interactions \cite{Mook:magnon_hall_effect_kagome:2014,Mook:topological_magnon_edge_state:2014,Mook:topological_magnon_kagome_interface:2015, owerre:magnon_haldane:2016,kim_et_al:magnon_haldane_kane_melel:2016} in our analysis; indeed, this would be straightforward and we could easily check whether the time-reversal-type symmetries $V_{1,3}^{\C}$ are preserved or broken. 

\section{Magnonic crystals} 
\label{magnonic_crystals}
When the spin wave is propagating in a magnonic crystal, \ie the magnetization $M_{\mathrm{s}}(x + \gamma) = M_{\mathrm{s}}(x)$ and exchange stiffness $A(x + \gamma) = A(x)$ are periodic with respect to some lattice $\Gamma$, spin waves can be efficiently expressed in terms of Bloch waves — just like the electronic wave function in condensed matter physics. Thanks to the Schrödinger form~\eqref{Krein_space_formalism:eqn:Schroedinger_form_spin_wave_equation} of the original linearized Landau-Lifshitz equation~\eqref{Krein_space_formalism:eqn:real_Schroedinger_Krein_form_linearized_LL_equations}, it is straightforward to go from intuition to mathematics. We continue to assume thermodynamic stability~\eqref{Krein_space_formalism:eqn:stability_condition}.

\subsection{Changing representation: the Bloch-Floquet-Zak transform} 
\label{magnonic_crystals:changing_representation}
Periodicity can be exploited with the exact same tools as in condensed matter physics: we employ a variant of the discrete Fourier transform, the (Bloch-Floquet-)Zak transform  \cite{Zak:dynamics_Bloch_electrons:1968}
\begin{align*}
	(\Fourier \Psi)(k,x) = \sum_{\gamma \in \Gamma} \e^{- \ii k \cdot (x + \gamma)} \, \Psi(x + \gamma) 
	, 
\end{align*}
to \emph{facilitate a change of representation;} compared to the usual Bloch-Floquet transform, it maps onto the space-periodic part of Bloch functions.

\subsubsection{The frequency band picture} 
\label{magnonic_crystals:changing_representation:frequency_band_picture}
The Zak transform decomposes the operator $H$ into a family $H(k) = W^{-1} \, \bigl ( D(k) + L(k) \bigr )$ depending on Bloch momentum $k$, where $W^{-1} = \sigma_2 \, M_{\mathrm{s}}$ is periodic and the other two operators 
\begin{subequations}\label{magnonic_crystal:eqn:D_L_operators_after_Bloch_Floquet}
	\begin{align}
		D(k) &= \mu_0 \, \gamma \, M_{\mathrm{s}}^{-1} \, \bigl ( H_0 + \nabla A \cdot \nabla M_{\mathrm{s}} + A \; \Delta M_{\mathrm{s}} \bigr ) 
		\, + \notag \\
		&\qquad 
		- \mu_0 \, \gamma \, \bigl ( \nabla A \cdot (\nabla - \ii k) + A \; (- \ii \nabla + k)^2 \bigr )
		, 
		\\
		L(k) &= - \ii \, \mu_0 \, \gamma \, \sigma_2 \, \biggl ( \sigma_3 \, (\partial_1 - \ii k_1) \, (\partial_2 - \ii k_2)
		\, + \biggr . 
		\notag 
		\\
		&\qquad \biggl . 
		+ \frac{\sigma_1 + \ii \sigma_2}{2} \, (\partial_2 - \ii k_2)^2 
		\, + \biggr . 
		\notag 
		\\
		&\qquad \biggl . 
		- \frac{\sigma_1 - \ii \sigma_2}{2} \, (\partial_1 - \ii k_1)^2 \biggr ) \; (- \ii \nabla + k)^{-2}
		, 
	\end{align}
\end{subequations}
can be computed explicitly. 

From the thermodynamic stability condition~\eqref{Krein_space_formalism:eqn:stability_condition} we deduce that $H(k)$ depends analytically on $k$; this means that near any point $k_0$ in the Brillouin zone the “hamiltonian” $H(k) = H(k_0) + \nabla_k H(k_0) \cdot (k - k_0) + \order \bigl ( (k - k_0)^2 \bigr )$ has a convergent Taylor expansion. Therefore, the frequency bands $\omega_n(k)$, which arise from the eigenvalue equation
\begin{align}
	H(k) \phi_n(k) &= \omega_n(k) \, \phi_n(k) 
	, 
	\label{magnonic_crystals:eqn:eigenvalue_equation_original_Krein_representation}
\end{align}
and after a suitable choice of phase the eigenfunctions $\phi_n(k)$ possess convergent Taylor expansions away from frequency band crossings. 

\subsubsection{Interplay of $\Fourier$ with the Krein structure} 
\label{magnonic_crystals:changing_representation:fourier_Krein}
As $W = \sigma_2 \, M_{\mathrm{s}}^{-1}$ is periodic, in addition to being unitary the Bloch-Floquet-Zak transform $\Fourier$ in fact defines a \emph{Krein}-unitary, 
\begin{align*}
	\sscpro{\Phi}{\Psi}_W = \bscpro{\Fourier \Phi}{\Fourier \Phi}_W 
	= \int_{\BZ} \dd k \, \bscpro{\Fourier \Phi(k)}{\Fourier \Phi(k)}_W
	, 
\end{align*}
where $\BZ$ is the first Brillouin zone and 
\begin{align*}
	\bscpro{\Fourier \Phi(k)}{\Fourier \Phi(k)}_W = \int_{\WS} \dd x \, \Fourier \Phi(k,x) \cdot \sigma_2 \, M_{\mathrm{s}}^{-1}(x) \, \Fourier \Phi(k,x)
\end{align*}
is a Krein inner product on the vector space $\Hil(M,\C^2)$ over the Wigner-Seitz cell $\WS$; we denote the resulting Krein space with $\Krein(M,\C^2) = \bigl ( \Hil(M,\C^2) , \scpro{\, \cdot \,}{\, \cdot \,}_W \bigr )$. 

Hence, $\Fourier \, H \, \Fourier^{-1}$ and $H(k) = H(k)^{\sharp}$ are Krein-hermitian operators. Moreover, the Krein-orthogonal projection associated to the frequency band $\omega_n(k)$ 
\begin{align}
	P_n(k) \psi(k) &= \sket{\phi_n(k)}_W \sbra{\phi_n(k)} \, \psi(k) 
	\notag 
	\\
	&
	= \sscpro{\phi_n(k)}{\psi(k)}_W \, \phi_n(k) 
	\label{magnonic_crystals:eqn:Krein_projection}
\end{align}
is expressed in terms of the “Krein-normalized” Bloch function $\phi_n(k)$ that satisfies $\sscpro{\phi_n(k)}{\phi_n(k)}_W = \mathrm{sgn} \, \omega_n(k) = \pm 1$: by thermodynamic stability, we have 
\begin{align*}
	0 < \bscpro{\phi_n(k)}{H_+(k) \phi_n(k)}_W 
	= \omega_n(k) \, \bscpro{\phi_n(k)}{\phi_n(k)}_W 
	, 
\end{align*}
and therefore the signs of $\omega_n(k)$ and $\bscpro{\phi_n(k)}{\phi_n(k)}_W \neq 0$ necessarily agree. Thus, we may normalize $\phi_n(k)$ to $\mathrm{sgn} \, \omega_n(k) = \pm 1$. 

Because Bloch functions are \emph{Krein}-orthonormal, this argument extends directly to linear combinations $\psi(k) = \sum_n \alpha_n(k) \, \phi_n(k)$ of positive frequency states. Also here, the Krein inner product $\scpro{\psi(k)}{\psi(k)}_W > 0$ will necessarily be positive as long as $\psi(k) \neq 0$ (see also \cite[Proposition~2]{Peano_Schulz_Baldes:topological_edge_states_bosonic_systems:2018}). 

When we sum up $P_n(k) = P_n(k)^{\sharp}$ for all positive/negative frequency bands, we obtain a Krein-orthogonal projection
\begin{align*}
	Q_{\pm}(k) &= \sum_{n : \pm \omega_n(k) > 0} P_n(k) 
\end{align*}
whose range is the positive/negative frequency subspace 
\begin{align}
	\Hil_{\pm}(k) = Q_{\pm}(k) \, \Krein(M,\C^2)
	. 
	\label{Krein_space_formalism:eqn:definition_Hil_pm}
\end{align}
Due to the thermodynamic stability assumption~\eqref{Krein_space_formalism:eqn:stability_condition} we know that $H(k)$ can be diagonalized with a Krein-unitary $U(k)$, 
\begin{align}
	\widetilde{H}(k) &= U(k) \, H(k) \, U(k)^{\sharp} = \widetilde{H}(k)^{\sharp} = \widetilde{H}(k)^{\dagger} 
	\label{magnonic_crystals:eqn:block_diagonalization_H_k}
	\\
	&= \left (
	\begin{matrix}
		h_+(k) & 0 \\
		0 & - \overline{h_+(-k)} \\
	\end{matrix}
	\right )
	, 
	\notag 
\end{align}
where the reality constraint $C \, H(k) \, C = - H(-k)$ in this representation translates to $C \, h_+(k) \, C = - \overline{h_+(-k)}$. Thus, \emph{frequency bands $\omega_{\pm n}(k) = - \omega_{\mp n}(-k)$ come in pairs} whose Bloch functions $\phi_{\pm n}(k) = \overline{\phi_{\mp n}(-k)}$ are related by complex conjugation. 

\subsubsection{Relation between $H(k)$ and $H_+(k)$} 
\label{magnonic_crystals:changing_representation:H_and_H_plus}
The notation already suggests that also for the $k$-dependent operators $H_+(k) = H(k) \, \vert_{\omega > 0}$ is the restriction to positive frequencies. When we restrict the Krein-unitary $U(k)$ to the $\omega > 0$ subspace $\Hil_+(k)$, we obtain another Krein unitary $U_+(k) = U(k) \, \vert_{\omega > 0}$ that relates $H_+(k)$ to 
\begin{align*}
	h_+(k) &= U_+(k) \, H_+(k) \, U_+(k)^{\sharp} 
\end{align*}
from Eq.~\eqref{magnonic_crystals:eqn:block_diagonalization_H_k} above. However, seeing as the sign of $\sscpro{\psi(k)}{\psi(k)}_W \geq 0$ is fixed for all positive frequency waves $\psi(k) \in \Hil_+(k)$, the Krein inner product $\scpro{\, \cdot \,}{\, \cdot \,}_W$ is also a \emph{scalar} product. Consequently, $H_+(k) = H_+(k)^{\sharp}$ and $h_+(k) = h_+(k)^{\sharp}$ are hermitian with respect to $\scpro{\, \cdot \,}{\, \cdot \,}_W$, and $U_+(k)$ is a $\scpro{\, \cdot \,}{\, \cdot \,}_W$-unitary. This unitary $U_+(k)$ maps Bloch functions $\phi_n(k)$ of $H_+(k)$ to Bloch functions $\varphi_n(k) = U_+(k) \, \phi_n(k)$ of $h_+(k)$ that satisfy the eigenvalue equation
\begin{align*}
	h_+(k) \varphi_n(k) &= \omega_n(k) \; \varphi_n(k) 
	. 
\end{align*}
Note that now $P_n(k) = P_n(k)^{\sharp}$ is an $\scpro{\, \cdot \,}{\, \cdot \,}_W$-orthogonal projection. Hence, expression~\eqref{intro:eqn:Chern_number} that involves the $\scpro{\, \cdot \,}{\, \cdot \,}_W$-orthogonal “Fermi projection” really yields an integer. 

\subsection{Existence of topologically protected magnonic edge modes via a magnonic bulk-edge correspondence} 
\label{magnonic_crystal:bulk_edge_correspondence}
Let us turn back to the problem that sparked our interest in the topic, Shindou et al.'s prediction of the existence of topologically protected edge modes in two-dimensional magnonic crystals. The theoretical basis for this prediction is the \emph{bulk-edge correspondence for class~A operators} (see \eg \cite[Chapter~7.2]{Prodan_Schulz_Baldes:complex_topological_insulators:2016}), that allows us to predict the net number of (left- vs.\ right-moving) edge modes solely from knowing the medium's bulk properties. This is embedded in the larger literature on topological insulators \cite{Altland_Zirnbauer:superconductors_symmetries:1997,Chiu_Teo_Schnyder_Ryu:classification_topological_insulators:2016,Kawabata_Shiozaki_Ueda_Sato:classification_non_hermitian_systems:2018}, that has been initially developed for quantum systems but has seen an increasing number of applications to classical waves (see \eg \cite{Raghu_Haldane:quantum_Hall_effect_photonic_crystals:2008,Wang_et_al:unidirectional_backscattering_photonic_crystal:2009,Ozawa_et_al:review_topological_photonics:2018,Suesstrunk_Huber:topological_phononic_edge_modes:2015,DeNittis_Lein:symmetries_electromagnetism:2017,DeNittis_Lein:equivalence_first_second_order_formalism_electromagnetism:2018}). 

Because we were able to frame the spin equation~\eqref{Krein_space_formalism:eqn:real_Schroedinger_Krein_form_linearized_LL_equations} in the form of the Schrödinger equation~\eqref{Krein_space_formalism:eqn:Schroedinger_form_spin_wave_equation}, all of the mathematical methods to analyze (quantum) topological insulators apply to spin waves as well. However, there may be important differences when it comes to the physical interpretation.

\subsubsection{Topological classification of the bulk} 
\label{magnonic_crystal:bulk_edge_correspondence:bulk_classification}
As we have shown in Section~\ref{Krein_space_formalism:topological_classification} the “hamiltonian” $H_+$ that enters the Schrödinger form~\eqref{Krein_space_formalism:eqn:Schroedinger_form_spin_wave_equation} of the classical spin equations~\eqref{Krein_space_formalism:eqn:real_Schroedinger_Krein_form_linearized_LL_equations} breaks time-reversal-type symmetry and due to the reality of spin waves $m(t) = \overline{m(t)}$ possesses no chiral-type or particle-hole-type symmetries. Therefore, in the parlance of the Ten Fold Way classification of topological insulators $H_+$ is in the \emph{same topological class as quantum hamiltonians that model systems exhibiting the Quantum Hall Effect — class~A} \cite{Altland_Zirnbauer:superconductors_symmetries:1997,Chiu_Teo_Schnyder_Ryu:classification_topological_insulators:2016,Prodan_Schulz_Baldes:complex_topological_insulators:2016}. 

Periodic class~A systems in two dimensions are classified with a single topological invariant, the well-known integer-valued Chern number~\eqref{intro:eqn:Chern_number}. Just like in solid state physics, if we are interested in the edge modes bridging the bulk band gap between the $n$th and $(n+1)$th frequency bands, the relevant Chern number $\mathrm{Ch}(P_{\mathrm{F}})$ is computed from the “Fermi projection”
\begin{align*}
	P_{\mathrm{F}}(k) = \sum_{j = 1}^n \sket{\phi_n(k)}_W \sbra{\phi_n(k)} 
\end{align*}
that includes contributions from the first $n$ frequency bands below the frequency band gap. Since $H_+(k)$ possesses no time-reversal-type symmetries, the Chern number $\mathrm{Ch}(P_{\mathrm{F}})$ can be different from zero. That is why it was important to identify the nature of $V_1^{\C}$ and $V_3^{\C}$ as potential time-reversal-type symmetries, and verify that the spin interactions included in Eq.~\eqref{Krein_space_formalism:eqn:real_Schroedinger_Krein_form_linearized_LL_equations} indeed break them. 

Lastly, we insist that $P_{\mathrm{F}}$ is \emph{not} to be interpreted as the physical state of the system, but enters the magnonic bulk-edge correspondence as an \emph{auxiliary quantity}. 

\subsubsection{Predicting the net number of edge modes via the class~A bulk-edge correspondence} 
\label{magnonic_crystal:bulk_edge_correspondence:bulk_edge_correspondence}
The concept of bulk-edge corresopndence is due to Hatsugai \cite{Hatsugai:Chern_number_edge_states:1993,Hatsugai:edge_states_Riemann_surface:1993}, and in this context reduces to two equalities: 
\begin{align}
	\mathrm{Ch}(P_{\mathrm{F}}) = T_{\mathrm{edge}} = \mbox{net \# of edge modes}
	\label{magnonic_crystal:eqn:bulk_edge_corresopndence}
\end{align}
The first one states that a topological invariant in the \emph{bulk} — the Chern number $\mathrm{Ch}(P_{\mathrm{F}})$ — equals a topological invariant $T_{\mathrm{edge}}$ defined on the \emph{edge}. This edge invariant can be computed in many different ways, \eg as a winding number of a Riemann surface \cite{Hatsugai:edge_states_Riemann_surface:1993}, a winding number defined from a smoothened version of $P_{\mathrm{F}}$ \cite[Eq.~(4.42)]{Prodan_Schulz_Baldes:complex_topological_insulators:2016} or as a \emph{spectral flow} \cite{Phillips:spectral_flow:1996,Booss_Bavnbek_Lesch_Phillips:spectral_flow:2002}. 

\begin{figure}
	\includegraphics[width=1.0\columnwidth]{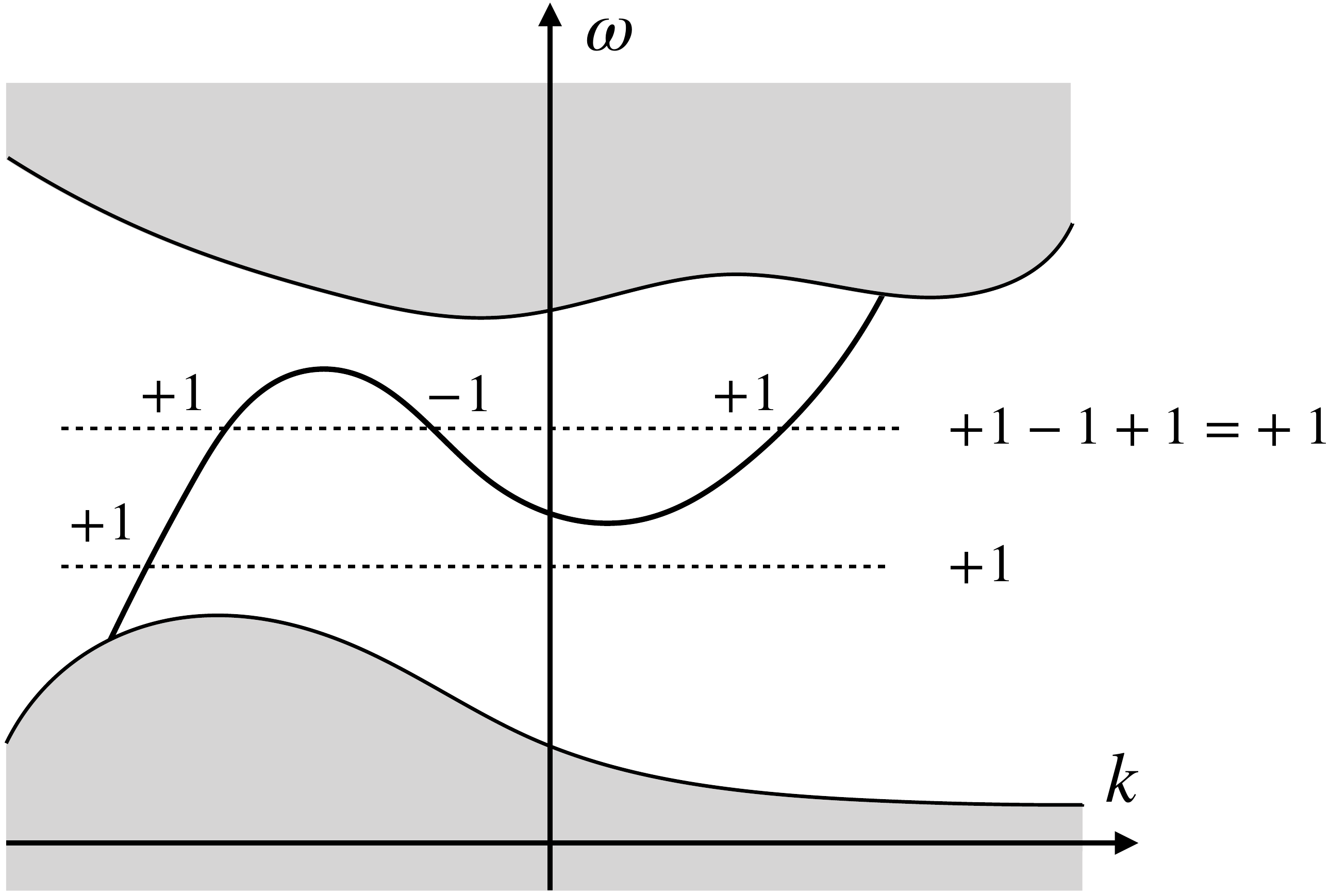}
	\caption{The spectral flow is the number of band intersections counted $\pm 1$ depending on whether the slope is positive or negative. It is independent of the choice of Fermi level — as long as it lies in the bulk band gap. Note that for classical waves we can create wave packets centered around a certain momentum and frequency, we can in fact distinguish the cases $+1 -1 + 1 = +1$ and $+1 = +1$ in experiment. However, only the net number of band intersections is topologically protected. }
	\label{magnonic_crystal:fig:edge_modes_spectral_flow}
\end{figure}

The last characterization makes the link to the physical observable, the net number of edge modes, obvious — which is the content of the second equality in Eq.~\eqref{magnonic_crystal:eqn:bulk_edge_corresopndence}. By definition the spectral flow \cite[Definition on p.~462]{Phillips:spectral_flow:1996} counts how often the edge bands intersect with the Fermi level from below (\ie with positive slope) minus the number of intersections from above (with negative slope). As the sign of the slope determines the edge mode's direction of propagation, this gives the net number of edge modes propagating left-to-right. We emphasize that only the \emph{net} number is topologically protected, not their total number or the number of left-/right-moving modes separately (\cf Figure~\ref{magnonic_crystal:fig:edge_modes_spectral_flow}). 

Note that in the Integer Quantum Hall Effect there is also a topological \emph{bulk} observable — the transverse bulk conductivity. But for spin waves there does not seem to be an associated bulk observable. 

To summarize, Shindou et al.'s prediction for the existence of topologically protected magnonic edge modes is footed on the bulk-edge correspondence~\eqref{magnonic_crystal:eqn:bulk_edge_corresopndence} for class~A operators. Mathematically speaking, that is all there is to it, Eq.~\eqref{magnonic_crystal:eqn:bulk_edge_corresopndence} holds no matter what waves these equations describe (\eg bosonic, fermionic or classical). On the other hand, it \emph{does} matter that the Krein structure plays no role at the end; for otherwise, we would have had to use a different classification scheme studied by De Nittis and Gomi \cite{DeNittis_Gomi:K_theoretic_classification_operators_on_Krein_spaces:2019}. And it \emph{is} crucial to insist on the real-valuedness constraint in order to eliminate unphysical symmetries. 

\section{Extension to the second-quantized equations} 
\label{second_quantized_equations}
So far our analysis concerned itself only with the \emph{classical} spin equation~\eqref{Krein_space_formalism:eqn:real_Schroedinger_Krein_form_linearized_LL_equations} rather than the more commonly studied second-quantized equations. However, we will show here that the second quantized equations inherits its topological classification from its classical counterpart. And consequently, our topological classification from Section~\ref{magnonic_crystal:bulk_edge_correspondence} directly applies to the second-quantized equations as well. 

Before second quantizing the linearization of Eq.~\eqref{Krein_space_formalism:eqn:nonlinear_Landau_Lifshitz_equations}, though, it is necessary to change coordinates: the Holstein-Primakoff transformation
\begin{align}
	T : \left (
	\begin{matrix}
		m_1(x) \\
		m_2(x) \\
	\end{matrix}
	\right ) \mapsto \left (
	\begin{matrix}
		m_-(x) \\
		m_+(x) \\
	\end{matrix}
	\right ) &= \frac{1}{\sqrt{2 M_{\mathrm{s}}(x)}} \left (
	\begin{matrix}
		m_1(x) - \ii m_2(x) \\
		m_1(x) + \ii m_2(x) \\
	\end{matrix}
	\right )
	\label{second_quantized_equations:eqn:Holstein_Primakoff_transformation}
\end{align}
normalizes the spin length and maps onto the eigenbasis of $\sigma_2$. 

The real-valuedness of the spin wave $m_j(x) = \overline{m_j(x)}$ manifests itself as $m_+(x) = \overline{m_-(x)}$ in Holstein-Primakoff coordinates. Formally treating $m_{\pm}$ as independent coordinates amounts to complexifying the equations; however, physically meaningful states are exactly those where $m_- = \overline{m_+}$. Translated to the level of operators, complex conjugation $C$ on the level of the original coordinates $(m_1,m_2)$ transforms to $T \, C \, T^{-1} = \sigma_1 \, C$ after Holstein-Primakoff transformation. Consequently, $\sigma_1 \, C$ is now the relevant particle-hole constraint which singles out real states. Because of this transformation the weight which enters the Krein inner product reduces from $W = \sigma_2 \, M_{\mathrm{s}}(x)^{-1}$ to merely $W' = (T^{-1})^{\dagger} \, W \, T^{-1} = \sigma_3$. 

After linearizing the Landau-Lifshitz equation~\eqref{Krein_space_formalism:eqn:nonlinear_Landau_Lifshitz_equations} in these new coordinates, $m_+$ and $m_-$ are promoted to creation and annihilation operators that satisfy bosonic commutation relations. 

Within the context of second quantization the Krein structure with weights $W' = \sigma_3$ is traditionally seen to emerge from the commutation relations of creation and annihilation operators — and \emph{Krein} unitaries \emph{preserve them} (\cf \cite[Section~4]{Colpa:diagonalization_boson_hamiltonian:1978}). However, the Krein structure also appears naturally in the classical context and stems from the symplectic structure (\cf Section~\ref{Krein_space_formalism:structure}). Therefore, it is not due to the quantum nature of Eq.~\eqref{second_quantized_equations:eqn:equations_of_motion_annihilation_creation_operators} below.

\subsection{Relation between classical and second-quantized equations} 
\label{second_quantized_equations:relation_to_classical_equation}
For simplicity, let us stay within the context of a periodic medium. The starting point of the second-quantized equations is a \emph{single-particle hamiltonian} $H_{\mathrm{sp}}(k)$, which enters as the “coefficient matrix” into the field hamiltonian 
\begin{align}
	\widehat{H}(k) = \frac{1}{2} \left (
	\begin{matrix}
		a(k) \\
		a^{\dagger}(-k) \\
	\end{matrix}
	\right )^{\dagger} \cdot H_{\mathrm{sp}}(k) \left (
	\begin{matrix}
		a(k) \\
		a^{\dagger}(k) \\
	\end{matrix}
	\right )
	. 
	\label{second_quantized_equations:eqn:second_quantized_hamiltonian}
\end{align}
Here, we have of course exploited periodicity via the Bloch-Floquet transform. $H_{\mathrm{sp}}(k)$ is assumed to be in one of the BdG classes, \ie it must possess a particle-hole-type “symmetry”. As before the root of this “symmetry” can be traced back to the real-valuedness of the spin waves. 

The single-particle hamiltonian is the link between the classical and the second-quantized equations: $H_{\mathrm{sp}}(k)$ is obtained from $D(k) + L(k)$ (given by Eq.~\eqref{magnonic_crystal:eqn:D_L_operators_after_Bloch_Floquet}) after applying the Holstein-Primakoff transformation~\eqref{second_quantized_equations:eqn:Holstein_Primakoff_transformation}. 

The Heisenberg equations of motion for annihilation and creation operators take the form of a Schrödinger equation on Krein space, 
\begin{align}
	\ii \frac{\dd}{\dd t} \left (
	\begin{matrix}
		a(k,t) \\
		a^{\dagger}(-k,t) \\
	\end{matrix}
	\right ) &= \left [ \hat{H} \, , \, \left (
	\begin{matrix}
		a(k,t) \\
		a^{\dagger}(-k,t) \\
	\end{matrix}
	\right ) \right ]
	\notag \\
	&= \sigma_3 \, H_{\mathrm{sp}}(k) \left (
	\begin{matrix}
		a(k,t) \\
		a^{\dagger}(-k,t) \\
	\end{matrix}
	\right )
	. 
	\label{second_quantized_equations:eqn:equations_of_motion_annihilation_creation_operators}
\end{align}
When the system is thermodynamically stable, then there exists a Krein unitary $U(k)$ which diagonalizes the single-particle hamiltonian to 
\begin{align}
	\widetilde{H}(k) &= U(k) \, \sigma_3 \, H_{\mathrm{sp}}(k) \, U(k)^{-1} 
	\notag \\
	&= \left (
	\begin{matrix}
		h_+(k) & 0 \\
		0 & h_-(k) \\
	\end{matrix}
	\right ) 
	= \left (
	\begin{matrix}
		h_+(k) & 0 \\
		0 & - \overline{h_+(-k)} \\
	\end{matrix}
	\right ) 
	\label{second_quantized_equations:eqn:Krein_unitary_transform_sigma_3_single_particle_hamiltonian}
\end{align}
also defines a new set of creation and annihilation operators $b_n^{\dagger}(k)$ and $b_n(k)$. Note that this operator conincides with the operator $\widetilde{H}(k)$ from Eq.~\eqref{magnonic_crystals:eqn:block_diagonalization_H_k} and therefore, 
\begin{align*}
	(b_n \psi)_{N-1}&(k_1 , \ldots , k_{N-1}) 
	= \\
	&= \sqrt{N} \, \int_{\BZ} \dd k \, \bscpro{\varphi_n(k) \; }{ \, \psi_N(k_1 , \ldots , k_{N-1} , k)}_{\WS}
\end{align*}
applies to an $N$-particle state destroys a Bloch wave $\varphi_n$ associated to the frequency band $\omega_n(k)$; note that the scalar product on the real space unit cell 
\begin{align*}
	 &\bscpro{\varphi_n(k) \; }{ \, \psi_N(k_1 , \ldots , k_{N-1} , k , x_1 , \ldots , x_{N-1})}_{\WS} = \\
	 &\quad 
	 = \int_{\WS} \dd x \, \overline{\varphi_n(k,x)} \; \psi_N(k_1 , \ldots , k_{N-1} , k , x_1 , \ldots , x_{N-1} , x)
\end{align*}
contains complex conjugation of its left argument $\varphi_n(k)$ and integrates out only one spatial variable. Similarly, the creation operator 
\begin{align*}
	(b_n^{\dagger} \psi)_{N+1}&(k_1 , \ldots , k_{N+1}) 
	= \\
	&= \frac{1}{\sqrt{N+1}} \sum_{j = 1}^{N+1} \varphi_n(k_j) \, \psi_N(k_1 , \ldots , \check{k}_j , \ldots , k_{N+1})
\end{align*}
creates a Bloch wave $\varphi_n$, where $\check{k}_j$ means this variable is omitted. Both of these operators act on the bosonic Fock space 
\begin{align*}
	\mathfrak{F} (\Fourier \Hil_+) = \bigoplus_{N = 0}^{\infty} \mathfrak{F}_N (\Fourier \Hil_+)
\end{align*}
that is the direct sum of the $n$-particle sectors 
\begin{align*}
	\mathfrak{F}_N (\Fourier \Hil_+) = \bigl ( \Fourier \Hil_+ \bigr )^{\otimes_{\mathrm{sym}} N}
	. 
\end{align*}
Each $n$-particle Fock sector is constructed by taking the $n$-fold symmetrized tensor product of the “classical” Hilbert space $\Hil_+(k)$ we have introduced in Eq.~\eqref{Krein_space_formalism:eqn:definition_Hil_pm}. The symmetrization encodes the bosonic nature of the waves so that \eg $\varphi_N(k_1,k_2,\ldots,x_1,x_2,\ldots) = \varphi_N(k_2,k_1,\ldots,x_2,x_1,\ldots)$ holds. By definition $\mathfrak{F}_0 = \C$ is the Fock vacuum. 

\subsection{Role of the particle-hole constraint in the second-order formalism} 
\label{second_quantized_equations:particle_hole_constraint}
On the level of the classical equations, we emphasized the distinction between a symmetry and a constraint: because physically meaningful spin vectors $m$ are real, on the level of the complexified equations complex conjugation becomes a constraint which singles out the physical states. 

The relation $m_- = C m_+ = \overline{m_+}$ translates to the fact that the creation operator is the adjoint of the annihilation operator. Put another way, the adjoint takes the place of complex conjugation; just like complex conjugation, the adjoint is \emph{anti}linear. This is how the redundancy emerges in the second-order formalism. 

The reason why we also need to flip the $k$ vector can also be understood from the classical equations: if we go from the position representation to the Bloch-Floquet representation, complex conjugation $C \Psi(x) = \overline{\Psi(x)}$ becomes $(\Fourier C \Psi)(k,x) = C (\Fourier \Psi)(-k,x)$. So in the Bloch-Floquet representation $2 \Re \Psi(x) = \Psi(x) + \overline{\Psi(x)}$ becomes $\Fourier \Psi(k,x) + \overline{\Fourier \Psi(-k,x)}$. 

Therefore, if we wish to make the real-valuedness constraint explicit in the second-order formalism in the Bloch-Floquet representation, we need to pair the annihilation operator $a(k)$ with the creation operator $a^{\dagger}(-k)$. 

The presence of the particle-hole constraint $\sigma_1 C$ in the single-particle hamiltonian is necessary to ensure that the time evolution preserves the relationship between the time-evolved creation operator $a^{\dagger}(t,-k) = \bigl ( a(t,k) \bigr )^{\dagger}$ and the annihilation operator. This becomes even more explicit when we instead work in the diagonalized basis where we compare $b_n(t,k) = \e^{- \ii t \omega_n(k)} \, b_n(k)$ with 
\begin{align*}
	\bigl ( b_n(t,k) \bigr )^{\dagger} = \e^{+ \ii t \omega_n(k)} \, \bigl ( b_n(k) \bigr )^{\dagger} 
	= \e^{- \ii t \omega_{-n}(-k)} \, b_n^{\dagger}(-k) 
	. 
\end{align*}
In the last equality we have exploited that due to the particle-hole constraint frequency bands come in pairs $\omega_{+n}(k) = - \omega_{-n}(-k)$. 

\subsection{Topological classification of the bulk} 
\label{second_quantized_equations:topological_classification}
Within the second-quantized framework the notion of topology stems from the single-particle hamiltonian $H_{\mathrm{sp}}$ via Eq.~\eqref{second_quantized_equations:eqn:Krein_unitary_transform_sigma_3_single_particle_hamiltonian}: the classification is via the Chern numbers computed from the Bloch functions of the positive frequency bands of $\sigma_3 \, H_{\mathrm{sp}}(k) = H(k)$, which are identical to those obtained from the classical hamiltonian $H(k)$ defined in Eq.~\eqref{Krein_space_formalism:eqn:contributions_spin_hamiltonian}. That is because the fundamental excitations of the second-quantized system are the eigenfunctions of $\sigma_3 \, H_{\mathrm{sp}}(k)$. Consequently, the topological properties of the classical system that $\sigma_3 \, H_{\mathrm{sp}}(k)$ describes are inherited by the second-quantized equations. 

As was the case for the classical equations, if the frequency bands $\omega_n$ and $\omega_{n+1}$ are separated by a gap, then the net number of gap traversing modes at the edge (taking the direction of travel into account) is given by the Chern number computed from the “Fermi projection” $P_{\mathrm{F}}(k) = \sum_{j = 1}^n \sket{\varphi_n(k)}_{W'} \sbra{\varphi_n(k)}$. 

While systems with even particle-hole-type symmetries (\eg class~D in dimension $1$ and $3$) support additional topological invariants in the form of a winding number, in the context of the second-order formalism of magnons complex conjugation does not play a role in the topological classification. So while the single-particle hamiltonian of the complexified equations is necessarily in one of the BdG classes, due to the real-valuedness of physical spin waves the particle-hole “symmetry” is in fact a constraint and does not matter for the topological classification of the \emph{physical system}. Hence, also here we are dealing with a hermitian system of class~A.%

\section{Extension to other thermodynamically stable bosonic BdG and certain classical wave equations} 
\label{other_systems}
Our strategy is not limited to magnons or the particular classical spin equations covered in Sections~\ref{Krein_space_formalism} and \ref{magnonic_crystals}. Rather, it applies to \emph{all} bosonic BdG equations~\eqref{intro:eqn:second_quantized_spin_equations} and classical equations
\begin{align}
	\ii \partial_t \psi(t) &= W \, D \psi(t)
	, 
	&&
	\psi(t_0) = m_0 = C m_0 \in \Hil
	, 
	\label{other_systems:eqn:classical_equations}
\end{align}
provided the following assumptions hold: 
\begin{enumerate}[(1)]
	\item $W = W^{\dagger}$ and $D = D^{\dagger}$ are hermitian operators on a Hilbert space $\Hil$, 
	\item $W$ is invertible, 
	\item $W = W^{\dagger}$ does not have a fixed sign ($\pm W \not > 0$), 
	\item $W \, D$ comes furnished with an even particle-hole-type symmetry $C$ (“complex conjugation”), and 
	\item $H = W \, D$ is thermodynamically stable, \ie $\pm \scpro{\psi}{D \psi} > 0$ holds. 
\end{enumerate}
Examples that we have not covered explicitly are magnons propagating in bilayer structures as studied by Katsura et al.\ \cite{Kondo_Akagi_Katsura:Z2_topological_invariant_magnons:2019}, spin waves in fermionic systems \cite{Mook:magnon_hall_effect_kagome:2014,Mook:topological_magnon_edge_state:2014,Mook:topological_magnon_kagome_interface:2015,owerre:magnon_haldane:2016,kim_et_al:magnon_haldane_kane_melel:2016} and phonons (see \eg \cite{Qin_Zhou_Shi:Berry_curvature_phonon_Hall_effect:2012}). 

As before, we begin with an analysis of the classical equations~\eqref{other_systems:eqn:classical_equations} and postpone the explanation as to how this applies to the corresponding second-quantized equation till Section~\ref{other_systems:quantum}.

\subsection{Certain classical wave equations} 
\label{other_systems:classical}
Many classical wave equations can be recast in a form that resembles~\eqref{other_systems:eqn:classical_equations}, although the precise conditions on $W$ may differ. Examples include Maxwell's equations in dielectrics and certain acoustic equations \cite{DeNittis_Lein:Schroedinger_formalism_classical_waves:2017}. The particle-hole \emph{constraint} that appears in the complexified equations stems from the fact that the original equations support real solutions; indeed, by fiat the actual waves are real. 

One of our motivations to pursue this work was to generalize the assumptions placed on $W$ in \cite{DeNittis_Lein:Schroedinger_formalism_classical_waves:2017}, where we have developed a Schrödinger formalism for certain classical waves. Specifically, we wanted to overcome the assumption that $\pm W > 0$ has a fixed sign and replace it with $W^{\dagger} W = W^2 > 0$ — as is the case when $W = \sigma_3 \otimes \id$. The mathematical challenges that stem from the fact that $\scpro{\phi}{\psi}_W = \scpro{\phi}{W \, \psi}$ is no longer a scalar, but only a \emph{Krein inner product} and Krein-hermiticity places much weaker constraints on an operator. For example, there need not be a resolution of the identity, \ie the operator may have Jordan blocks. 

This Krein structure emerges not just in the context of magnons, but for many other wave equations. One notable example are simple models for the propagation of electromagnetic waves in metals (see \eg \cite{Bliokh_Leykam_Lein_Nori:topological_classification_homogeneous_electromagnetic_media:2019}): here the electric permittivity $\eps < 0$ which enters the “material weights” $W = 
\left ( 
\begin{smallmatrix}
	\eps & 0 \\
	0 & \mu \\
\end{smallmatrix}
\right )$ is negative whereas $\mu > 0$. The other operator is the \emph{free Maxwell operator} $D = 
\left ( 
\begin{smallmatrix}
	0 & + \ii \nabla^{\times} \\
	- \ii \nabla^{\times} & 0 \\
\end{smallmatrix}
\right ) = - \sigma_2 \otimes \nabla^{\times}$ that contains the curl $\nabla^{\times} \mathbf{E} = \nabla \times \mathbf{E}$ in its offdiagonal. 

To exclude technical complications, we needed to impose Assumption~(5), thermodynamic stability~\eqref{Krein_space_formalism:eqn:stability_condition}. Ideally, we would like to drop this assumption as it excludes Maxwell's equations for metals ($D = - \sigma_2 \times \nabla^{\times}$ has no fixed sign), and replace it with a weaker condition. However, at present this is beyond reach. 

Nevertheless, in case the thermodynamic stability condition our discussion from the end of Section~\ref{Krein_space_formalism:structure:primer} applies verbatim: $H = H^{\sharp} = W^{-1} \, H^{\dagger} \, W$ is a Krein-hermitian operator that is connected to a hermitian operator $\widetilde{H}$ by a similarity transform facilitated by a Krein-unitary. This operator $\widetilde{H}
= \left ( 
\begin{smallmatrix}
	h_+ & 0 \\
	0 & - C \, h_+ \, C \\
\end{smallmatrix}
\right )
$ is block-diagonal and the two block operators $h_- = - C \, h_+ \, C$ are related by complex conjugation. The even particle-hole-type “symmetry” $K = \sigma_1 \otimes C$ of the complexified equations is purely block-offdiagonal and can be traced back to the fact that \eqref{other_systems:eqn:classical_equations} supports real solutions, \ie 
\begin{align*}
	\e^{- \ii t H} \, C = C \, \e^{- \ii t H}
	\; \; \Longleftrightarrow \; \; 
	\e^{- \ii t \widetilde{H}} \, K = K \, \e^{- \ii t \widetilde{H}}
	. 
\end{align*}
This allows us to implement the strategy outlined in Section~\ref{Krein_space_formalism:Schroedinger_formalism:reduction_to_complex_omega_0_spin_waves} and get rid of redundant fields in the complexified equations: we can uniquely represent real fields $m = C m = 2 \Re \psi$ as the real part of a complex $\omega > 0$ wave $\psi = Q_+ m$, where $Q_+$ is the Krein-hermitian projection onto the positive frequencies defined through Eq.~\eqref{Krein_space_formalism:eqn:positive_frequency_Krein_projection}. 

Just like in Section~\ref{Krein_space_formalism:Schroedinger_formalism:K_plus_Hilbert_space} we again see that on the positive frequency subspace $\Krein_+$, the Krein inner product $\scpro{\, \cdot \,}{\, \cdot \,}_W$ is positive definite — and hence, a \emph{scalar} product. That means on this subspace $\Krein_+ = \Hil_+$, the operator $H_+ = H \vert_{\omega > 0} = H_+^{\sharp}$ is hermitian in the usual sense. Overall, this shows that classical wave equation~\eqref{other_systems:eqn:classical_equations} is equivalent to 
\begin{align*}
	\ii \partial_t \psi(t) = H_+ \psi(t) 
	,
	&&
	\psi(t_0) = Q_+ m_0 \in \Hil_+ 
	. 
\end{align*}
Put succinctly, under these five conditions we can \emph{equivalently} recast the classical wave equation~\eqref{other_systems:eqn:classical_equations} to an equation that takes, mathematically speaking, the form of a Schrödinger equation. The hermiticity of the generator of the dynamics $H_+$ allows us to adapt methods initially developed for quantum systems to that classical wave equation. 

\subsection{Symmetries vs.\ constraints} 
\label{other_systems:symmetries_vs_constraints}
\emph{Any equivalent formulation} for a given system contains the exact same phenomenology and necessarily has to have the exact same characteristics. That applies to symmetries, too: any symmetry of the physical system must manifest itself in any of the formulations. But independently of the mathematical formulation, all of these formulations need to have the same number and type of symmetries. And consequently, the topological classification of all of these different equations needs to be consistent. 

The central point is to distinguish between actual symmetries of the system and constraints, which appear to be symmetries if we forget about the real-valuedness of the physical fields. How to do this was described in Section~\ref{Krein_space_formalism:Schroedinger_formalism:symmetries}: in the complexified formalism, we need to deal with Krein-(anti)unitaries $U$ which are block-diagonal with respect to the positive/negative frequency splitting; thus, $U_+$ restricts to a proper (anti)unitary on the positive frequency subspace $\Hil_+$. This rules out complex conjugation as that maps positive onto negative frequency states, and is therefore block-offdiagonal in the positive/negative frequency splitting; equivalently, the Krein-(anti)unitary must commute with $H$. This \emph{selection rule} tells us how to implement a symmetry of the real waves on the complex Hilbert space $\Hil_+$: we can either implement a symmetry of the real vectors (Krein-)unitarily or (Krein-)antiunitarily. 

\subsection{Second-quantized bosonic BdG systems} 
\label{other_systems:quantum}
Very often the starting point is the second-quantized equations~\eqref{intro:eqn:second_quantized_spin_equations} where spin waves can be excited and destroyed. Here, the weights $W = \sigma_3 \otimes \id$ stem from the commutation relations of creation and annihilation operators and $D = H_{\mathrm{sp}}$ is the single-particle hamiltonian. To ensure that the time-evolved annihilation and creation operators are adjoints of one another, which can be viewed as a constraint, $H_{\mathrm{sp}}$ needs to come with a particle-hole-type “symmetry”. 

The second-quantized equations give rise to an associated classical equation of the form~\eqref{other_systems:eqn:classical_equations} that comes furnished with a particle-hole-type constraint. Following the arguments in Section~\ref{second_quantized_equations}, we conclude that the second-quantized equation inherits the topological classification from the associated classical equation. 

\subsection{Topological classification} 
\label{other_systems:topological_classification}
The distinction between symmetries and constraints leads to subtle, but essential differences in the topological classification: while it is true that bosonic BdG systems of the form~\eqref{intro:eqn:second_quantized_spin_equations} are defined in terms of an operator with a particle-hole-type symmetry, we need to keep in mind that all physically meaningful states of this equation are necessarily particle-hole symmetric. Back in the classical context, if the state can be described by \eg a projection $P$, then $P$ needs to commute with $C$. Moreover, because the whole physical system — states, observables and dynamics — can be described solely on the level of the complex Hilbert space $\Hil_+$, we deduce that $P$ needs to commute with $H$ and therefore the projection $P = P_+ + P_-$ cleanly splits into positive/negative frequency parts $P_{\pm} = Q_{\pm} \, P \, Q_{\pm} = P \, Q_{\pm}$. This is quite different from, say, the Dirac equation where particles and anti-particles can be excited selectively and independently from one another. Here, pure states represented by the projection $P$ need not commute with the particle-hole symmetry. 

Moreover, there is also a second, more implicit fact in play here: in conventional quantum systems with a particle-hole symmetry, the relevant band gap is located at $E = 0$, whereas this is \emph{not} the case for bosonic BdG and classical wave equations. Instead, we are interested in frequency gaps for some $\omega_0 > 0$. 

Because in our context only commuting symmetries matter, barring any crystallographic symmetries or so (which could lead to an analog of the valley Hall effect, see \eg \cite{Wu_Hu:topological_photonic_crystal_with_time_reversal_symmetry:2015}), the equations are classified either as class~A (no symmetries), class~AI (presence of an even time-reversal-type symmetry) or class~AII (presence of an odd time-reversal-type symmetry as in \cite{Kondo_Akagi_Katsura:Z2_topological_invariant_magnons:2019}). However, we could of course apply the recently obtained classification of topological crystalline insulators \cite{Shiozaki_Sato_Gomi:band_topology_3d_crystallographic_groups:2018} to the systems discussed here. 

For example, in case the system is periodic and of class~A, our analysis in Section~\ref{magnonic_crystal:bulk_edge_correspondence} proves we have a bulk classification in terms of Chern numbers, which in turn explains the presence of topologically protected edge modes in 2d systems via the bulk-edge correspondence~\eqref{magnonic_crystal:eqn:bulk_edge_corresopndence}. 

Lastly, we would like to contrast and compare our approach with a work by \cite{Kennedy_Zirnbauer:Bott_periodicity_Z2_symmetric_ground_states:2016}: they classified \emph{fermionic} BdG systems under the assumption that states below the Fermi energy are filled, which leads to the constraint $C \, P \, C = \id - P$ where $P = 1_{(-\infty,0]}(H)$. This constraint is equivalent to ours above. And therefore, fermionic class~D systems with this constraint on the state are in fact equivalent to a system of class~A (see \cite[Section~2.1]{Kennedy_Zirnbauer:Bott_periodicity_Z2_symmetric_ground_states:2016}). While their results do not apply \emph{directly} to bosonic BdG systems, Kennedy and Zirnbauer place emphasis on the exact same point: it is \emph{not enough} to merely study the hamiltonian, we need to include conditions that single out physically relevant and meaningful states in our analysis — which could translate to constraints in the topological classification. 

\section{Conclusion} 
\label{conclusion}
To summarize our findings, we have derived the topological classification of second-quantized bosonic BdG equations~\eqref{intro:eqn:second_quantized_spin_equations} and associated classical wave equations~\eqref{intro:eqn:classical_wave_equation}. In particular our results include equations that describe classical and quantum spin excitations in suitable media. For the equations considered by Shindou et al.\ \cite{Shindou_et_al:chiral_magnonic_edge_modes:2013} we conclude that what they propose is indeed an analog of the Quantum Hall Effect in the following precise sense: the operators which describe the two systems are of class~A, and are therefore classified up to gap-preserving deformations by Chern numbers. The commonly studied second-quantized equations inherit the topological classification from their classical counterpart. 

To perform the initial steps of our analysis, rewriting the classical spin equations in the form of a Schrödinger equation, taking the Krein structure into account was crucial. The Krein structure is not particular to the second-quantized equations, though, where it is attributed to the bosonic nature of creation and annihilation operators. In case of the classical equations, the Krein structure emerges from the symplectic structure of phase space. 

We have shown how to implement physical symmetries in the Schrödinger formalism. In principle, our arguments can be applied to crystalline and other symmetries by combining our work with \eg \cite{Shiozaki_Sato_Gomi:band_topology_3d_crystallographic_groups:2018}. 

We reckon that our results will be helpful to understand other classical waves better whose description also involves a Krein structure; examples are electromagnetic media whose material weights are not positive (\eg when the medium is metallic \cite{Bliokh_Leykam_Lein_Nori:topological_classification_homogeneous_electromagnetic_media:2019}) or certain acoustic metamaterials \cite{Hussein_Leamy_Ruzzene:dynamics_phononic_materials_review:2014,Ma_Sheng:acoustic_metamaterials:2016}. However, in these cases the “thermodynamic stability” condition~\eqref{Krein_space_formalism:eqn:stability_condition} is violated, and consequently, the operators involved will be more complicated. 

For example, we expect that performing a topological classification of such operators is a lot more involved as we will have to apply more sophisticated topological classification schemes for non-hermitian systems; recent research by Kawabata et al.\ \cite{Kawabata_Shiozaki_Ueda_Sato:classification_non_hermitian_systems:2018} suggests there are 38 distinct topological classes. Luckily, we could avoid such complications here — just: while the original operator is not hermitian, thanks to the thermodynamic stability condition, it can be transformed into one. Therefore, the usual, well-established theory suffices to understand bosonic BdG systems and their classical counterparts. 

\section*{Acknowledgements} 
\label{acknowledgements}
The authors would like to thank Kei Yamamoto for many enjoyable discussions on this topic, which kickstarted this project. Both authors have been supported by JSPS (grant numbers 16K17761 and JP17H06460, respectively) and by a Fusion Grant from the WPI-AIMR. Moreover, we would like to thank Hosho Katsura for friendly discussions on an early draft of this work. 

\bibliographystyle{apsrev4-1}
\bibliography{bibliography}

\end{document}